\documentclass[11pt,a4paper]{article}
\usepackage{times}
\usepackage{tgtermes}
\usepackage[T1]{fontenc}
\usepackage[utf8]{inputenc}
\usepackage[french]{babel}
\usepackage{fancyhdr}
\usepackage{setspace}
\usepackage{comment}

\usepackage{graphicx}
\usepackage{lipsum}
\usepackage{array}
\usepackage{caption}
\usepackage{multicol}
\usepackage{hyperref}
\usepackage{afterpage}
\usepackage{setspace}
\usepackage{pgffor}
\usepackage{amsmath}
\usepackage{amsmath,amssymb,bm}
\usepackage{graphicx}
\usepackage{tikz,graphics,color,fullpage,float,epsf,caption,subcaption}

\usepackage{titlesec}
\titleformat{\section}
  {\normalfont\fontsize{16}{20}\bfseries}{\thesection}{1em}{}
\titleformat{\subsection}
  {\normalfont\fontsize{16}{20}\bfseries}{\thesubsection}{1em}{}
\titlespacing*{\section}
{0pt}{2.ex plus 1ex minus .2ex}{.0ex}
\titlespacing*{\subsection}
{0pt}{0.ex}{.0ex}

\begin{document}

\begin{center}
\begin{spacing}{2.05}
{\fontsize{18}{18} \selectfont
\bf
Variational Approach to Viscoelastic Fracture: Comparison of a phase-field and of a lip-field approach
}
\end{spacing}
\end{center}
\vspace{-1.25cm}
\begin{center}
{\fontsize{12}{14} \selectfont
\bf
Rajasekar Gopalsamy\textsuperscript{a}, Nicolas Chevaugeon\textsuperscript{b}, Olivier Chupin\textsuperscript{a}, Ferhat Hammoum\textsuperscript{c} \\
\bigskip
}
\end{center}
{\fontsize{10}{10} \selectfont
a. Univ Gustave Eiffel, MAST-LAMES, F-44344 Bouguenais, France \\
b. Ecole Centrale de Nantes, GeM Institute, UMR CNRS 6183,1 rue de la Noe,
44321 Nantes, France\\
c. Univ Gustave Eiffel, MAST-MIT, F-44344 Bouguenais, France
}

\vspace{10pt}

{\fontsize{16}{20}
\bf
Abstract :
}
\bigskip

\textit{Fracture of viscoelastic materials is considered to be a complex phenomenon due to their highly rate sensitive behavior. In this context, we are interested in the quasi-static response of a viscoelastic solid subjected to damage. This paper outlines a new incremental variational based approach and its computational implementation to model damage in viscoelastic solids.  The variational formalism allows us to embed the local constitutive equations into a global incremental potential, the minimization of which provides the solution to the mechanical problem. Softening damage models in their local form are known to result in spurious mesh-sensitive results, and hence non-locality (or regularization) has to be introduced to preserve the mathematical relevance of the problem. In the present paper, we consider two different regularization techniques for the viscoelastic damage model: a particular phase-field and a lip-field approach. The model parameters are calibrated to obtain some equivalence between both these approaches. Numerical results are then presented for the bidimensional case and both  these approaches compare well. Numerical results also demonstrate the ability of the model to qualitatively represent the typical rate-dependent behaviour of the viscoelastic materials. Besides, the novelty of the present work lies in the use of lip-field approach for the first time in a viscoelastic context.}

\vspace{28pt}
{\fontsize{14}{20}
\bf
Key words :}  
{\fontsize{14}{20}damage, viscoelasticity, lip-field, phase-field, variational approach
}

\section{Introduction}
\medskip
Viscoelastic materials like asphalt, biological tissues, wood and polymers have numerous applications in engineering.  The properties of these materials are highly rate and temperature dependent, therefore, greatly affecting their failure behavior \cite{Lakes,Amico}. Designing for the mitigation of fractures in viscoelastic materials is an important problem, for example, in the case of asphalt and pavement construction \cite{Pirmohammad}. 

Fracture mechanics commenced with the earlier work of Griffith \cite{Griffith}.  Based on this theory, the stress at the crack tip becomes infinite. Fracture mechanics is used to characterise the loads on crack using a single parameter. A number of different parameters have been developed and used. When the non-linear zone also called the Fracture Process Zone (FPZ) (which involves cavity formation, stringing, chain pull-outs for polymers, and bond breaking) is relatively small compared to the crack size, the cracking can then be described by the elastic forces within the bulk of the material and is termed as Linear Elastic Fracture Mechanics (LEFM). The stress state at the crack tip can then be characterised by the Stress Intensity Factor (SIF). The analytical expression for SIF for several configurations is available in the literature \cite{Anderson}.  The crack is considered to grow rapidly when the SIF reaches the critical SIF (or fracture toughness), which is usually determined from experiments. 
However, since all the materials available in nature contain FPZ of finite width, the SIF approach cannot be used in most cases, as the crack growth is significantly affected by the non-linear process occurring in the vicinity of the crack tip.  
Hence, a different parameter used for characterising crack growth is based on the energy, called the energy release rate $G$. This approach is called the energy approach and crack growth is considered to occur when the energy available for crack growth is sufficient to overcome the resistance to fracture of the material. Griffith was the first to propose the energy criterion for fracture \cite{Griffith} . The criterion for crack propagation is given as follows: $G \geq Gc$. $Gc$ is the critical energy release rate. This energy criterion was ignored for a long period until the early 1950s, since the Griffith derivation for the analytical expression of $G$ was available only for the linear elastic fracture case (or brittle fracture). Irwin was the pioneer in developing the present version of this approach through a modified energy criterion that accounts for the non-linear effects in the FPZ through the inclusion of the other (usually plastic) dissipation term \cite{Irwin}. Several approaches have been introduced since then to compute the energy release rate. J-integral is the most common way to compute the energy release rate \cite{Rice}.

Methods developed for the linear elastic and elastic-plastic fracture cannot be used for the time dependent viscoelastic fracture mainly due to the viscous dissipation in the bulk causing time dependent fracture. Knauss \cite{Knauss} presents a detailed review on time-dependent viscoelastic fracture. For viscoelastic problems without fracture, it is usually common practice to find the solution by use of the correspondence principle \cite{Brinson}. This principle allows to find the viscoelastic solution by solving for an associated elastic problem in the transform domain (using Laplace Transform) followed by inverting the elastic solution back to the time domain. For fracture in non-linear viscoelastic material under large strain, Schapery makes use of the correspondence principle to define a generalized time-dependent J-integral analogous to the non-linear elastic case \cite{Schapery 1984} \cite{Schapery 1990}.  Based on this theory, a new model to study viscoelastic fracture under creep loading has been proposed in \cite{Dubois}. This model extends the initial Griffith energy balancefor linear elastic solids to viscoelasticity by inclusion of the viscous dissipation term, and crack growth is driven only by the free energy.  In the mean time, X-FEM \cite{Belytschko, Moes} has gained a lot of attention. X-FEM approach was used in \cite{Zhang 2009, Zhang 2010} to address linear viscoelasticity problems with inclusions and cracks. However, the Griffith-type models have certain drawbacks. a) They are not able to predict crack initiation and can only be dealt in the case of pre-existing cracks (initial cracks).  b) They are not able to predict crack branching. The Griffith type models were later enhanced with damage mechanics based approaches. This can be broadly categorised as Cohesive Zone Model (CZM) \cite{Dugdale1, Barenblatt}  and Continuum Damage Mechanics (CDM) \cite{Kachanov,Lemaitre}. Both approaches introduce a length scale into the model accounting for the micro-damage before fracture that can be associated to the nucleation, coalescence and growth of voids to form macro-cracks.  Some works on the use of CZM to model viscoelastic fracture can be found in \cite{Yoon, Rahulkumar, Buttlar}. In the later, a bilinear CZM is used to simulate the viscoelastic fracture behavior of asphalt concrete. However it was also reported that the numerical simulations were found to diverge when the orientation of cohesive elements is not tailored to the crack path observed from experiments. CDM is yet another popular approach to model damage in viscoelastic materials. Some of the commonly used CDM approaches are non-local integral damage models \cite{Pijaudier, Lorentz}, kinematic \cite{Schreyer} and damage based \cite{Nguyen1,Peerlings1} gradient models,  phase-field damage models \cite{Miehe, Ambati}  and Thick Level Set(TLS) approach \cite{Moes TLS}.   For example, Shiferaw et al. \cite{Benjamin} used the TLS approach to model damage in viscoelastic materials for a wide range of temperatures. In the later, the authors also demonstrated the ability of the model to represent experimental results for the uni-axial tension test under monotonous loading rate. Phase-field models have also been used to model cracks in viscoelasticity \cite{Ambati visco,Shen, Yin}. For e.g. in \cite{Shen} both free energy and a part of viscous dissipation is used to drive the phase damage. Another class of methods that was recently developed under the formalism of CDM was the Lip-field approach \cite{Moes LF, Chevaugeon} to fracture. The main difference of this approach from the phase-field approach lies in the fact that the damage dissipation potential is not a function of the damage gradient. 

Fracture behavior of viscoelastic materials is a complex phenomenon due to their highly rate-sensitive behavior. This is mainly attributed to the viscoelastic dissipation in the bulk material around the crack tip and inertia effects (for high speed cracks) contributing to the fracture toughness. Modeling such fracture behavior can still be considered to be a relatively open area. This sets the objective of the present paper. In the present work, we limit ourselves the quasi-static crack growth.  We consider the formalism of generalized standard materials \cite{Germain,Halphen}, where we define the free energy potential, the viscous dissipation potential, and the damage dissipation potential (all convex). The constitutive laws (or evolution equations) are then prescribed using the relation between the driving forces (thermodynamic forces), free energies, and dissipation potentials.   An incremental potential is then defined in line with the works of \cite{Stainier,Fancello, Lahellec}  on variationally consistent incremental principles for dissipative systems. By variationally consistent, we mean that the problem can be formulated as an optimisation problem with respect  to the fields of state variables.  They also posses attractive features that they offer the possibilities for extensive mathematical and numerical analysis \cite{Radovitzky}. This incremental formalism also embodies in itself the fully implicit time discretisation. The formalism of generalized standard materials ensures that incremental potential is convex with respect to each state variable separately. Hence, the classical alternate minimisation, where minimization over each state variable while freezing other variables could be used to find the solution. Moreover, in the present work, only the free energy is used to drive damage growth. This helps in preserving energy conservation in the bulk.  Variational consistency also allows one to state the problem in an elegant and compact format.

The undamaged free energy and viscous dissipation potentials corresponding to the Generalized Kelvin-Voigt (GKV) model \cite{Brinson} have been considered to represent the viscoelastic behavior. We are also interested in comparing the results for the lip-field and phase-field regularization techniques. The difference lies in the definitions for the damage dissipation potential: a local damage dissipation potential corresponding to lip-field and a non-local (gradient damage type) corresponding to phase-field. The local expression of dissipation potentials is well known to trigger spurious localization \cite{Milan1}. Hence in case of lip-field, an external regularization is introduced through the addition of a new space called the Lipschitz space (a non-local space) and the search for damage field is constrained to lie in this space. This helps in preserving the ellipticity of the problem. The bounds estimate proved in \cite{Moes LF} also helps to greatly simplify the computational process of constraining the damage field to lie in Lipschitz space. Besides, in order to aid the comparison of the results corresponding to lip-field and phase-field approaches, some equivalence is derived between the respective model parameters by equating the fracture energies.

The paper is structured as follows.  Initially, the variational framework for damage in viscoelasticity (quasi-static case) is discussed in the following section. In this section, the damage dissipation potentials of both phase-field and  lip-field approaches are considered. In section \ref{sec:eq}, an equivalence is derived between phase-field and lip-field approaches to aid in their comparison. Following the theory, the details of the implementation are presented in Section \ref{sec:ca}. Section \ref{sec:sr} presents the results of the finite element analysis carried out on some benchmark tests. Finally, section \ref{sec:cn} concludes the paper.

\section{The viscoelastic damage model: variational framework}

\medskip
In this section, the fully implicit time discretized incremental potential for rate-dependent systems, associated to the problem of damage in viscoelasticity is derived. This allows to reformulate the problem as an optimisation problem.

Consider the deformation of a body initially occupying a domain $\Omega$ through a displacement field $\mathbf{u}$. We assume small, isotropic and quasi-static deformations under isothermal conditions. The Cauchy stress and small strain tensor is denoted as $ \bm{\sigma}$ and $ \bm{\varepsilon}$.
\begin{equation}
     \bm{\varepsilon(u)} = \frac{1}{2} (\nabla \mathbf{u} + \nabla^T \mathbf{u}) \label{eq:eq1}
\end{equation}
Regarding boundary conditions, displacement is applied to a part of the boundary $\Gamma_u$. We assume zero traction forces on the rest of the boundary $\Gamma_N$ and that there is no body force (without loss of generality). To be kinematically admissible at any given time $t$, the displacement field $\mathbf{u}$ must belong to $U(t)$:
\begin{align}
U(t) = \{\mathbf{u} \in H^1(\Omega) : \mathbf{u} = \mathbf{u_d}(t)\;\; on \;\;\Gamma_u\}
\label{eq:eq2}
\end{align}
The equilibrium condition then reads
\begin{align}
    \int_{\Omega}  \bm{\sigma} :  \bm{\varepsilon}(\mathbf{u^*}) \; d\Omega = 0, \;\;\; \forall \; \mathbf{u^*}\in U^*\;\;\;\; U^* =\{\mathbf{u} \in H^1(\Omega) : \mathbf{u} = 0\;\; on \;\;\Gamma_u\}
    \label{eq:eq3}
\end{align}

The equilibrium and the kinematic equations Eq. (\ref{eq:eq3}) and (\ref{eq:eq2}) must be complemented by the constitutive equations.

A spectral approach is used to describe the viscoelastic behaviour where the set of internal variables are strain based. The internal state of the viscoelastic material is then described by the following state variables: $ \bm{\varepsilon},  \bm{\varepsilon_1},..., \bm{\varepsilon_n}$ and $d$, where $ \bm{\varepsilon_1},.., \bm{\varepsilon_n}  $ are the internal (or viscous) strains and $d$ is the scalar damage variable. In the rest of the work $ \bm{\varepsilon_i}$ will be used to indicate the set of internal strains $\{ \bm{\varepsilon_1},.., \bm{\varepsilon_n} \}$. The formalism of generalized standard materials is followed, where we consider two convex thermodynamic potentials: the free energy potential $\psi$ and the dissipation potential $\phi$. 
\begin{align}
    \psi &= \psi ( \bm{\varepsilon}, \bm{\varepsilon_i},d)  \label{eq:eq4}\\
    \phi  &=  \phi_v(  \bm{\dot{\varepsilon}_i}) + \phi_d (d, \dot{d})   \label{eq:eq5}
\end{align}
where the additive decomposition of the dissipation potential $\phi$ into the viscous dissipation potential $\phi_v$ and the damage dissipation potential $\phi_d$ (both convex) has been used. The viscous dissipation potential is not considered a function of the scalar damage variable. The consequence is that only free energy is used to drive damage and also helps in preserving energy conservation.

The driving forces conjugate to state variables are then given by 
\begin{align}
     \bm{\sigma} = \frac{\partial \psi}{\partial \bm{\varepsilon}}  \hspace{1cm}
     \bm{\sigma_{\eta,i}} = -\frac{\partial \psi}{\partial \bm{\varepsilon_i}}   = \frac{\partial \phi_v}{ \partial \bm{\dot{{\varepsilon}_i}}} \hspace{1cm}
    Y = -\frac{\partial \psi}{\partial d}  = \frac{\partial \phi_d}{\partial \dot{d}} \; when \; \{\dot{d} > 0\;  \&\; d<1\} \label{eq:eq6}
\end{align}
where $\sigma$ is the stress, $\sigma_{\eta,i}$ is the viscous stress conjugate to viscous strain $ \bm{\varepsilon_i}$ and $Y$ is the local energy release rate conjugate to $d$.  The negative sign in viscous stress and local energy release rate indicates the dissipative nature of the respective driving forces. {In Eq. (\ref{eq:eq6}.3) , equality  is guaranteed when the $\dot{d} > 0$  and $d < 1$} (damage criterion in strict sense). 

The time domain of study $[0,T]$ is discretized into time intervals $t_0, t_1,...,t_m,t_{m+1},...t_M=T$, where $t_{m+1}-t_m =t-t_m = \Delta t$, where we drop the $m+1$ index for simplicity and the same will be adopted for all the state variables for the rest of the paper. Hence $\{ \bm{\varepsilon_m},  \bm{\varepsilon_{i,m}},d\}$ and  $\{ \bm{\varepsilon},  \bm{\varepsilon_i}, d\}$ are used to indicate the state of the material at time steps $t_m$ and $t_{m+1}=t$.  An implicit Euler time discretisation has been used to discretize the time derivative of internal strains ($ \bm{\dot{\varepsilon}_{i,m+1}} =  \bm{\dot{\varepsilon}_{i}} =  ( \bm{{\varepsilon}_{i}} -  \bm{{\varepsilon}_{i,m}}) / \Delta t$ ). Eq. (\ref{eq:eq6}) then yields the following constitutive equation in differential form at the time step $t_{m+1} = t$
\begin{align}
    & \bm{\sigma} =  \frac{\partial \psi}{\partial \bm{\varepsilon}} ( \bm{\varepsilon},  \bm{\varepsilon_{i}}, d)   \hspace{.2cm}
     &\frac{\partial \psi}{\partial \bm{\varepsilon_i}} ( \bm{\varepsilon},  \bm{\varepsilon_{i}}, d)  + \frac{\partial \phi_v}{ \partial \bm{\dot{\varepsilon}_i}} \left(\frac{ \bm{{\varepsilon}_{i}}- \bm{{\varepsilon}_{i,m}}}{\Delta t}\right) = 0 \hspace{.45cm}
    &\mu = \frac{\partial \psi}{\partial d} ( \bm{\varepsilon},  \bm{\varepsilon_{i}}, d) + \frac{\partial \phi_d}{\partial \dot{d}} (d, \dot{d})  \label{eq:eq7}
\end{align}
where $\mu$ is a scalar variable associated to damage criterion given by the Karush-Kahn-Trucker (KKT) conditions
\begin{align}
    &\mu - \lambda_1 + \lambda_2 = 0 \label{eq:eq8} \\
    &\lambda_1 \geq 0, \hspace{.5cm} d-d_m \geq 0,\hspace{.5cm}  \lambda_1(d-d_m) = 0  \label{eq:eq9}\\
    &\lambda_2 \geq 0,\hspace{.5cm} 1-d \geq 0,\hspace{.8cm} \lambda_2(1-d) = 0 \label{eq:eq10}
\end{align}
$\lambda_1$ and $\lambda_2$ are the Lagrange multipliers associated with irreversibility of the damage ($\dot{d} \geq 0)$ and the bound constraints $(d\leq 1)$. Eq. (\ref{eq:eq7}) along with the KKT conditions ( Eq. (\ref{eq:eq8}-\ref{eq:eq10})) are the constitutive equations describing the material behaviour.

By defining a new potential  $\Tilde{\phi}_d$ (also convex) associated to damage as follows,
\begin{align}
    \frac{\partial \Tilde{\phi}_d}{\partial d} = \frac{\partial \phi_d}{\partial \dot{d}} \label{eq:eq11}
\end{align}
It can be readily seen that the local evolution equations for internal strains and damage ( Eq. (\ref{eq:eq7}.2), (\ref{eq:eq8})) are recovered as the stationarity conditions  of the following variational problem.
\begin{align}
    &\underset{\lambda_1, \lambda_2}{\max}\;  \;\underset{ \bm{\varepsilon_i , d}}{ {\min}} \hspace{.2cm} J \hspace{1cm} \text{such\; that\;} \lambda_1\geq0,\; \lambda_2 \geq 0\\
    &J = \psi( \bm{\varepsilon},  \bm{\varepsilon_{i}}, d) + \Delta t \phi_v \left(\frac{ \bm{{\varepsilon}_{i}}- \bm{{\varepsilon}_{i,m}}}{\Delta t} \right) + \Tilde{\phi}_d (d, \dot{d}) - \lambda_1 (d-d_m) - \lambda_2 (1-d)   \label{eq:eq12}
\end{align}
The above potential also allows to rewrite Eq. (\ref{eq:eq7}.1) as 
\begin{align}
     \bm{\sigma} = \frac{\partial J}{\partial  \bm{\varepsilon}} \label{eq:eq13}
\end{align}
To this end, the problem of finding the internal state of the material at a given time step $t$ can be reformulated into an optimisation problem as follows: 
\begin{align}
    \{ \mathbf{u},  \bm{\varepsilon_i}, d \}=  \arg  \underset{d' \in A_m}{\underset{ \bm{\varepsilon_i}' \in P}{\underset{\mathbf{u}' \in U_m  }{  \min}}}  F(\mathbf{u}',  \bm{\varepsilon_i}',d'\;;\;\mathbf{u_m},  \bm{\varepsilon_{i,m}},d_m, \Delta t )  \label{eq:eq14}
\end{align}
where the global incremental potential $F$ is given as follows
\begin{align}
    F =\int_{\Omega} f \;d \Omega  =\int_{\Omega} \psi( \bm{\varepsilon},  \bm{\varepsilon_{i}}, d) + \Delta t \phi_v (\frac{ \bm{{\varepsilon}_{i}}- \bm{{\varepsilon}_{i,m}}}{\Delta t}) + \Tilde{\phi}_d (d, \dot{d})   d \Omega  \label{eq:eq15}
\end{align}
$J$ is the Lagrangian function associated with $f$. Note that the minimization of $F$ w.r.t. $\mathbf{u}$ results in the equilibrium equation Eq. (\ref{eq:eq3}).  The admissible spaces for the state variables are given follows: 
\begin{align}
    &U_m = U(t_{m+1}) = U(t)  \label{eq:eq16}\\
    &P = \{ \mathbf{q} : \mathbf{q} \in L^{\infty}(\Omega)\}  \label{eq:eq17}\\
    &A_m = \{ d \in L^{\infty}(\Omega): d_m \leq d \leq 1 \} \label{eq:eq18}
\end{align}

where $U(t)$ is given by Eq. (\ref{eq:eq2}) and the space $A_m$ ensures the damage irreversibility and bound constraints.

\begin{figure}[H]
\centering
{\includegraphics[trim= 25  600  25 19, width= 12cm, height=5cm]{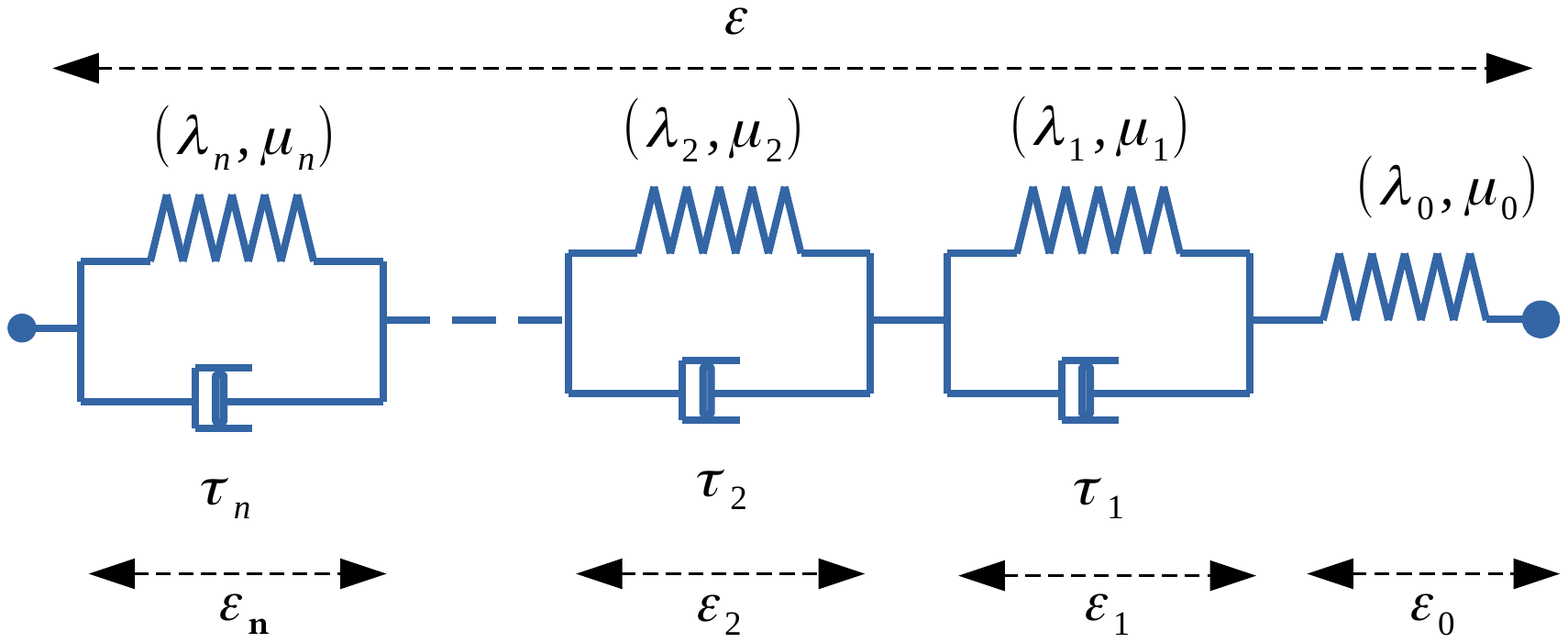}}
   \caption{Generalized Kelvin Voigt (GKV) model with $n$ Kelvin Voigt (KV) units }
   \label{fig:1} 
\end{figure}

Having laid the basis for the variational framework, the equations can be closed by defining the free energy potential  $\psi$ and the dissipation potential $\phi$. In this work, we consider the Generalized Kelvin-Voigt (GKV) model to describe the viscoelastic behavior. The schematic of GKV for one dimensional case (without damage) is shown in Figure \ref{fig:1} .
$\{(\lambda_0, \mu_0),..,(\lambda_n, \mu_n)\}$ and $\{\tau_1,..,\tau_n\}$ are the Lame's constants and retardation times of each Kelvin-Voigt (KV) unit respectively. The total strain can then be written as
\begin{align}
     \bm{\varepsilon} = \sum_{i=0}^{n}  \bm{\varepsilon_i} \label{eq:eq19}
\end{align}
where $ \bm{\varepsilon_0}$ is the strain in the free spring. This model allows to define $\psi$ and $\phi_v$, while the choice of $\phi_d$ depends on the choice of regularization (lip-field or phase-field).
For the GKV model considered, the free energy and viscous dissipation potential are given as follows:
\begin{align}
    &\psi = \sum_{i=0}^{n} \left(
    g(d) \mu_i < \bm{\varepsilon_i}>_+ \mathbf{:} < \bm{\varepsilon_i}>_+  + g(\beta d) \mu_i < \bm{\varepsilon_i}>_- \mathbf{:} < \bm{\varepsilon_i}>_-  + g(d) \frac{\lambda_i}{2}  Tr( \bm{\varepsilon_i})^2
    \right)  \label{eq:eq20}\\
     &\phi_v = \sum_{i=1}^n  \tau_i  \left( \mu_i  \bm{\dot{\varepsilon}_i} \mathbf{:}  \bm{\dot{\varepsilon}_i} + \frac{\lambda_i}{2}  Tr( \bm{\dot{\varepsilon}_i})^2 \right) \label{eq:eq21}
\end{align}
where the definitions $< \bm{\varepsilon_i}>_+$ and $< \bm{\varepsilon_i}>_-$ are given in Appendix \ref{ap:a}. $0 \leq \beta \leq 1$ is a user-defined parameter that controls behavior in tension and compression. If $\beta = 1$, the behavior of the material is symmetric in tension and compression. If $\beta =0$, the material is asymmetric in the sense that it is able to store some portion of free energy in compression (and hence recover some stiffness in compression). In the present work, we use the following commonly used choice for $g(d)$. ($g(0)= 1 \; g(1) = 0$ and $g(d)$ convex on $[0,1]$).
\begin{align}
    g(d) = (1-d)^2   \label{eq:eq22}
\end{align}
For the symmetrical case ($\beta = 1$), Eq. (\ref{eq:eq20}) becomes
\begin{align}
    \psi = \sum_{i=0}^{n} \left(
    g(d) \mu_i  \bm{\varepsilon_i} \mathbf{:}  \bm{\varepsilon_i}  + g(d) \frac{\lambda_i}{2}  Tr( \bm{\varepsilon_i})^2
    \right) = g(d) \psi^+   \label{eq:eq23}
\end{align}
For the asymmetric case ($\beta=0$), Eq. (\ref{eq:eq20}) becomes
\begin{align}
    \psi =  g(d)\; \psi^+ + \psi^-   \label{eq:eq24}
\end{align}

where $\psi^+$ is the part of the free energy that that drives damage growth and $\psi^-$ is the portion of free energy not affected by damage. 

The minimization of $F$ w.r.t. $d$ results in spurious mesh dependent results. This can be overcome by regularization techniques that introduce length scale(s) into the model. In the next sections, we look briefly at the phase-field and the lip-field regularization. It should be noted that both the phase-field and lip-field damage fields are represented by the same scalar damage variable $d$ and if any misconception state is identified, it will be further clarified with more details.

\subsection{Phase-field regularization}
In this case, the introduction of regularizing length scale $l_1$ is made through the definition of the following dissipation potential.
\begin{align}
    \phi_d = \frac{G_c}{4l_1}\; \delta_d(h(d) + 2l_1^2 |\nabla d|^2) \;\dot{d} \hspace{.5cm} \implies \hspace{.5cm} \Tilde{\phi}_d = \frac{G_c}{4l_1}\; (h(d) + 2l_1^2 |\nabla d|^2) = G_c \gamma (d, \nabla d)  \label{eq:eq25}
\end{align}
where $\delta_d$ is used to represent the functional derivative w.r.t damage variable $d$. The damage gradient term ensures the non-locality of damage field. $G_c$ is the fracture toughness in the sense of Griffith ($J/m^2$). $h(d)$ is the damage softening function and $\gamma(d, \nabla d)$ is called the crack surface functional in the phase-field community and in the global sense, the later gives a measure of crack surface area in three-dimensions (crack length in 2D). The relation between $\phi_d$ and $\Tilde{\phi}_d$ is given by Eq. (\ref{eq:eq11}). With the definition of the non-local phase-field dissipation potential, the stationarity problem given by Eq. (\ref{eq:eq14}) can be re-written as follows:
\begin{align}
     \{ \mathbf{u},  \bm{\varepsilon_i}, d \}=  \arg \underset{d' \in R_m }{\underset{ \bm{\varepsilon_i}' \in P}{\underset{\mathbf{u}' \in U_m  }{  \min}}}  F(\mathbf{u}',  \bm{\varepsilon_i}',d'\;;\;\mathbf{u_m},  \bm{\varepsilon_{i,m}},d_m, \Delta t )   \label{eq:eq26}
\end{align}
where $R_m \subset A_m$ is defined as
\begin{align}
    R_m = \{ d \in H^{1}(\Omega): d_m \leq d \leq 1 \}  \label{eq:eq27}
\end{align}
The increased regularity of the space $R_m$ over which the phase-damage field is sought is due to the presence of gradient term in damage dissipation potential.

\subsection{Lip-field regularization}
In contrast to non-local dissipation potential considered for phase-field, the dissipation potential in local form is considered for the lip-field approach as follows
\begin{align}
    \phi_d =  Y_c h'(d) \dot{d}  \hspace{1cm} \implies \hspace{1cm}   \Tilde{\phi}_d = Y_c h(d)   \label{eq:eq28}
\end{align}
$Y_c$ is the critical energy release rate per unit volume ($J/m^3$). The non-locality is then through the introduction of Lipschitz space $\mathcal{L}_{\Omega}$ and by constraining the damage field to lie in this non-local space.
\begin{align}
    \mathcal{L}_{\Omega} = \left \{ d \in L^{\infty}(\Omega) : | d(\mathbf{x}) - d(\mathbf{y}) | \leq \frac{1}{l_2} dist(\mathbf{x},\mathbf{y}) \;\; \forall  \mathbf{x},\mathbf{y} \in \Omega  \right \}  \label{eq:eq29}
\end{align}
where $l_2$ is the lip-field regularizing length scale parameter. The bound $1/l_2$ for the magnitude of slope is also called the \textit{Lipschitz constant}.  The introduction of new space allows us to reformulate the variational problem given by Eq. (\ref{eq:eq14}) as follows.
\begin{align}
\{\mathbf{u},  \bm{\varepsilon_i}, d \}=  \arg \underset{d' \in A_m \cap \mathcal{L}_{\Omega}}{\underset{ \bm{\varepsilon_i}' \in P}{\underset{\mathbf{u}' \in U_m  }{  \min}}}  F(\mathbf{u}',  \bm{\varepsilon_i}',d'\;;\;\mathbf{u_m},  \bm{\varepsilon_{i,m}},d_m, \Delta t )  \label{eq:eq30}
\end{align}
where $d' \in \mathcal{A}_m \cap \mathcal{L}_{\Omega}$ ensures irreversibility and non-locality of damage field.

\section{Model parameters calibration}
\label{sec:eq}
In order to aid the comparison of phase-field and lip-field approaches for viscoelastic fracture, some equivalence between them are derived in this section. To be precise, the relation between the regularizing length scales and fracture parameters of both the approaches are derived. For the rest of the study, we consider the following damage softening function:
\begin{align}
    h(d) = 2d^2  \label{eq:eq31}
\end{align}
 For the considered quadratic damage softening function, the damage starts to grow at the onset of loading for both phase-field \cite{Bourdin2007, Charlotte} and lip-field \cite{Moes LF} models. However, its often used in phase-field community due to its relative ease of implementation \cite{Azinpour}. In case of lip-field approach \cite{Chevaugeon}, a different (convex) softening function doesn't alter the numerical implementation.    

\subsubsection{Phase-field like damage shape function}
\begin{figure}[H]
  \centering
  {\includegraphics[trim=1cm 23.5cm 1cm 2cm,clip,scale= .8]{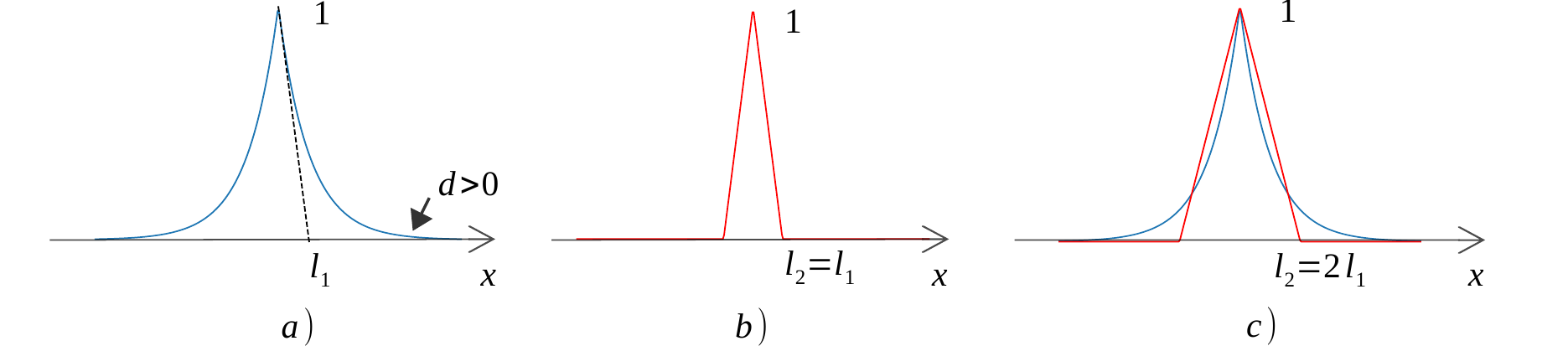}}
  \caption{damage profiles of phase-damage (blue) and lip-damage (red) : a) phase-damage profile with length scale $l_1$ b) lip-damage profile with length scale $l_2 = l_1$ and c) phase-damage profile with length $l_1$ and lip-damage profile for $l_2 = 2l_1$ . }\label{fig:2}
\end{figure}

The relation between the length scales $l_1$ and $l_2$ is derived in this section aiding to have a similar damage shape function for both the phase-field and lip-field approaches. The damage profile of the phase-field models is known a priori \cite{Miehe} and is given by Eq. (\ref{eq:eq32}) for the considered phase-field model with crack located at centre $x=0$ (1D).
\begin{align}
    d= e^{-\frac{|x|}{l_1}}  \label{eq:eq32}
\end{align}
In case of lip-field, the damage profile is not known a priori. But numerical results for brittle and quasi-brittle fracture \cite{Moes LF, Chevaugeon} indicate that lip-field damage profile tries to saturate the Lipschitz constraint with the magnitude of the slope of damage reaching the upper bound given by the Lipschitz constant. Hence we safely assume the damage profile to be linear in this case and is given by Eq. (\ref{eq:eq33}) for a crack located at $x=0$ (1D).
\begin{align}
d = \left\{
                \begin{array}{ll}
                  1-|x|/l_2 \; &\forall \;|x| \leq l_2\\
                  0  &elsewhere
                \end{array}
              \right\}  \label{eq:eq33}
\end{align}
The exponential and the assumed linear damage profile for the phase-field and lip-field for $l_1 = l_2$ are shown in Figures \ref{fig:2}.a and \ref{fig:2}.b. It can be seen that in case of phase-field, $d \rightarrow 0$ only when $x \rightarrow \infty$. 
$l_1 = l_2$ is not considered as the best choice (see Figures \ref{fig:2}.a and \ref{fig:2}.b), as they result in different damage distribution around the crack. The Lipschitz constant for the considered phase-field damage profile is given by $1/ l_1$. Due to the exponential profile of phase-field, the Lipschitz constant of lip-field is set by averaging the maximum (Lipschitz constant of phase-field) and minimum ($0\; as \;x \rightarrow \infty$) slope of phase-field.
\begin{align}
    \frac{1}{l_2} = \frac{\frac{1}{l_1} + 0} {2}   \hspace{.5cm} \implies \hspace{.5cm} l_2 = 2l_1  \label{eq:eq34}
\end{align}
The damage profile in this case is plotted in Figure \ref{fig:2}.c and the difference in distribution of phase-field and lip-field damage profiles is observed to be relatively less.

\subsubsection{Fracture energy}
Having obtained some equivalence in damage profiles for phase-field and lip-field models, the relation between the fracture energy parameters $Y_c$ and $G_c$ is sought in this section. In order for both the models to be energetically equivalent, the damage energy per unit of crack advance has to be the same.  In case of phase-field, fracture energy per unit of crack advance is directly given by the parameter $G_c$ , while in case of lip-field, for the considered damage profile Eq. (\ref{eq:eq33}), the fracture energy per unit of crack length for a fully developed crack is given by $\int_{-l_2}^{l_2} \Tilde{\phi}_d (d(x)) dx$.  For energetically equivalent, the following relation holds
\begin{align}
    \int_{-l_2}^{l_2} Y_c h(d(x)) dx = G_c \hspace{.5cm} \implies \hspace{.5cm} Y_c = \frac{3 G_c}{4 l_2 }  \label{eq:eq35}
\end{align}
The relations given by Eq.(\ref{eq:eq34}) and (\ref{eq:eq35}), allows us to compare both the regularization approaches, provided the fracture parameters for any one of these models are given.

\section{Computational aspects}
\label{sec:ca}
The function $F$ is convex w.r.t. $\{\mathbf{u}, \bm{\varepsilon_i}\}$ and $\{d\}$ separately and the admissible spaces for the state variables are also convex. Hence an alternate minimization (or staggered scheme) is used, resulting in a  series of convex optimization problem. At time step $t_{m+1} = t$, the following iterations are repeated until convergence.
\begin{enumerate}
    \item Find $\{ \mathbf{u}^{k+1},  \bm{\varepsilon_i}^{k+1}\}$ with fixed $d = d^{k}$
    \item Find $d^{k+1}$ with fixed $\{\mathbf{u}^{k+1},  \bm{\varepsilon_i}^{k+1}\}$
\end{enumerate}
where $k$ represents the alternate minimisation iteration number.

\subsection{Finding displacements and internal strains}
\textit{Case A:} For the symmetric case ($\beta=1$), the minimisation of $F$ w.r.t. $\{\mathbf{u}, \bm{\varepsilon_i}\}$ results in a linear problem. In this section, only the final equations are presented and the details could be found in the Appendix \ref{ap:b1}. The problem of finding displacements is given by the following weak form.
\begin{align}
    \int_{\Omega} { \bm{\varepsilon}}(\mathbf{u^*}) : \mathbb{H} : { \bm{\varepsilon}}(\mathbf{u})\; d \Omega = \int_\omega   { \bm{\sigma_{int}}} :{ \bm{\varepsilon}}(\mathbf{u^*}) \; d \Omega \hspace{.5cm} \forall \;\mathbf{u^*}\in U^*\;, \mathbf{u}\in U_m  \label{eq:eq36}
\end{align}
 The space $U_m$ is given by Eq. (\ref{eq:eq16}) and the space for virtual displacements $\mathbf{u^*}$ is given in Eq. (\ref{eq:eq3}). The expressions for the tangent matrix $\mathbb{H}$, internal stress ${ \bm{\sigma_{int}}}$ are provided in the Appendix \ref{ap:b1}. It should be noted that, the weak form given by Eq. (\ref{eq:eq36}) is similar to elasticity with a modified tangent matrix and an additional body force term ($\bm{\sigma_{int}}$) accounting for viscoelasticity. The body force term depends on damage at current time ($d^k$) and internal strains from previous time step ($\bm{\varepsilon_{i,m}}$).

Having found the displacements and strain, the internal strains are updated using  Eq. (\ref{eq:eq69}) and Eq. (\ref{eq:eq68}).

\textit{Case B:} For the asymmetrical case ($\beta=0$).  The minimization of $F$ w.r.t. $\{\mathbf{u}, \bm{\varepsilon_i}\}$ results in the following set of non-linear equations
\begin{align}
    &\int_{\Omega}  \bm{\sigma} (  \bm{\varepsilon} (\mathbf{u}),  \bm{\varepsilon_i}, d) :  \bm{\varepsilon} (\mathbf{u^*}) \;d \Omega=0  \label{eq:eq38}\\
    &\int_{\Omega} \frac{\partial \psi }{\partial  \bm{\varepsilon_i}} (  \bm{\varepsilon}(\mathbf{u}),  \bm{\varepsilon_i},d) + \Delta t \frac{\partial \phi_v}{\partial  \bm{\varepsilon_i}} ( \bm{\varepsilon_i},  \bm{\varepsilon_{i,m}}) \; d \Omega =0 \hspace{.5cm} \forall i \in \{1,2,..,n\} \label{eq:eq39}
\end{align}
where $\mathbf{u^*}\in U^*, \mathbf{u}\in U_m$ and $ \bm{\varepsilon_i} \in P(\Omega)$. The non-linearity come from the eigen split used to express the free energy. These $n+1$ non-linear equations are solved using the Newton-Raphson method (see Appendix \ref{ap:b2}).

\subsection{Phase-field damage solver}
For the considered $g(d)$ and $h(d)$, we obtain the AT2 (Ambrosio-Tortorelli-2) implementation of the phase-field model  \cite{Bourdin2007, Miehe,Miehe2}. The damage is then given as follows:
\begin{align}
    d = arg \min_{d' \in R_m} F(\mathbf{u},  \bm{\varepsilon_i},d', \nabla d')   \label{eq:eq40}
\end{align}
Though the direct optimisation of Eq. (\ref{eq:eq40}) is possible over the entire domain $\Omega$ (for example, using Matlab Optimization toolbox \cite{Amor} or the active set Newton method \cite{Benson} to impose the constraints), it is considered very expensive because of the non-locality of $d$. Hence in this work, we employ the commonly used local history field variable \cite{Miehe} to impose the irreversibility constraints.

For the AT2 model, $0\leq d\leq 1$ is intrinsically guaranteed (see Eq. (\ref{eq:eq32})). However, the irreversibility constraint $\dot{d} \geq 0$ (or $d > d_m$) has to be enforced, so that $d$ lies in the admissible space $R_m$.  The minimization problem could then be written through the set of Karush–Kuhn–Tucker (KKT) conditions
\begin{align}
    \frac{\delta F}{ \delta d} \geq 0 \hspace{1cm} \dot{d} \geq 0 \hspace{1cm} \frac{\delta F}{ \delta d} \dot{d} = 0  \label{eq:eq41}
\end{align}
where $\delta F / \delta d$ is the variational derivative of $F$ w.r.t. $d$. We follow the use of a local history variable \cite{Miehe2} to enforce irreversibility constraint. This allows us to replace the the variational inequality in Eq. (\ref{eq:eq41}) by variational equality as follows.
\begin{align}
    \frac{\delta F}{\delta d} = 0, \hspace{1cm} F = \int_\Omega g(d) H + \psi^-  +\Delta t \phi_v \left(\frac{ \bm{{\varepsilon}_{i}}- \bm{{\varepsilon}_{i,m}}}{\Delta t} \right) +  G_c \gamma (d, \nabla d) \; d \Omega  \label{eq:eq42}
\end{align}
with $H$ defined as follows
\begin{align}
    H(\mathbf{x},t) = \max_{s \in [0,t] } \psi^+ (\mathbf{x},s)  \label{eq:eq43}
\end{align}
In this case, Eq. (\ref{eq:eq42}) results in the following strong form for phase-damage $d$
\begin{align}
    G_c \left( \frac{1}{l_1} d - l_1 \Delta d \right) - 2(1-d) H = 0 \hspace{1cm} \Vec{\nabla d} . \Vec{n} = 0 \;\;on \;\; \partial \Omega \label{eq:eq44}
\end{align}
The associated weak from can be written as follows: 
\begin{align}
    \int_{\Omega} G_c \left( \frac{1}{l_1} d \;\delta d + l_1 \nabla d. \nabla \delta d \right) - 2(1-d) \delta d \;H  \;\; d \Omega = 0 \hspace{1cm} \forall\; d, \delta d \in S(\Omega)   \label{eq:eq45}
\end{align}
The space $S(\Omega) $ is given by
\begin{align}
    S(\Omega) = \{d : d \in H^1(\Omega) \}  \label{eq:eq46}
\end{align}
The use of history variable allows one to search for damage field in $S(\Omega) \subset R_m(\Omega)$. However, the consequence is the loss in variational structure of the problem leading to damage bound constraint not being enforced in a strong sense.

\subsection{Lip-field damage solver}
\label{sec:lf}
In contrast to the previous case, the  direct optimization to find the lip-damage is considered to be relatively efficient, thanks to the bounds estimation and its properties proposed in \cite{Moes LF}.  For any given damage field $d^*$, the upper $\overline{d}(\mathbf{x})$ and lower $\underline{d}(\mathbf{x})$ bounds are defined as
\begin{align}
    \Bar{d}(\mathbf{x}) = \max_{\mathbf{y} \in \Omega} ( d^* (\mathbf{y}) -\frac{1\;}{l_2} dist(\mathbf{x},\mathbf{y})) \hspace{.8cm}
    \underline{d}(\mathbf{x})  = \min_{\mathbf{y} \in \Omega} ( d^* (\mathbf{y}) +\frac{1\;}{l_2} dist(\mathbf{x},\mathbf{y})) \hspace{.8cm} \forall \mathbf{x} \in \Omega \label{eq:eq47}
\end{align}
The steps involved in finding damage at any given alternate minimization iteration can be listed as follows
\begin{enumerate}
    \item Find the local damage field $d_{loc}$ defined as follows 
    \begin{align}
        d_{loc} (\mathbf{x}) = arg \min_{d' \in A_m} F(\mathbf{u},  \bm{\varepsilon_i},d')  \hspace{.3cm} \forall \; \mathbf{x} \in \Omega  \label{eq:eq48}
    \end{align}
    
    \item Compute the bounds from Eq. (\ref{eq:eq47}) for  $d^* = d_{loc}$. These bounds have been proved to satisfy the following properties.
    \begin{align}
        &d_m(\mathbf{x}) \leq \underline{d}(\mathbf{x}) \leq d_{loc}(\mathbf{x}) \leq \Bar{d}(\mathbf{x}) \leq 1  \label{eq:eq49}\\ &\underline{d}(\mathbf{x})\leq d(\mathbf{x}) \leq \Bar{d}(\mathbf{x}) \label{eq:eq50} \\ &\overline{d}(\mathbf{x})-\underline{d}(\mathbf{x}) = 0 \implies d_{loc}(\mathbf{x})=d(\mathbf{x}) \in \mathcal{L}_{\Omega} \label{eq:eq51}
    \end{align} 
    Hence in the region where the bounds are equal (also called the \textit{inactive zone}), the local damage solution $d_{loc}$ is same as the sought optimal (lip-)damage field $d$. 
    
    \item The problem yet to be solved is to find damage in the active zone $\overline{\Omega}$ which is defined as follows : 
    \begin{align}
        \overline{\Omega} = \{\mathbf{x} \in \Omega: \underline{d} (\mathbf{x}) \neq \overline{d}(\mathbf{x}) \}  \label{eq:eq52}
    \end{align}
    The damage field in the active zone can then be found by the following optimisation
    \begin{align}
        d = arg \min_{d' \in \overline{A}_m \cap \mathcal{L}_{\overline{\Omega}}} F_{\;\overline{\Omega}}( \bm{\varepsilon}(\mathbf{u}),  \bm{\varepsilon_i}, d'), \hspace{1cm} F_{\;\overline{\Omega}} = \int_{\overline{\Omega}} f d\overline{\Omega}  \label{eq:eq53}
    \end{align}
    where the space $\overline{A}_m$ is defined as follows.
    \begin{align}
    \overline{A}_m = \{ d \in L^{\infty}(\overline{\Omega}): \underline{d} \leq d \leq \overline{d} \}  \label{eq:eq54}
    \end{align}
\end{enumerate}
The use of bounds to simplify the projection of a given damage field into Lipschitz space is demonstrated with an example in the Appendix \ref{ap:c}.

\subsection{Spatial discretisation and lip constraints }
\begin{figure}[H]
\centering
\includegraphics[scale = .3]{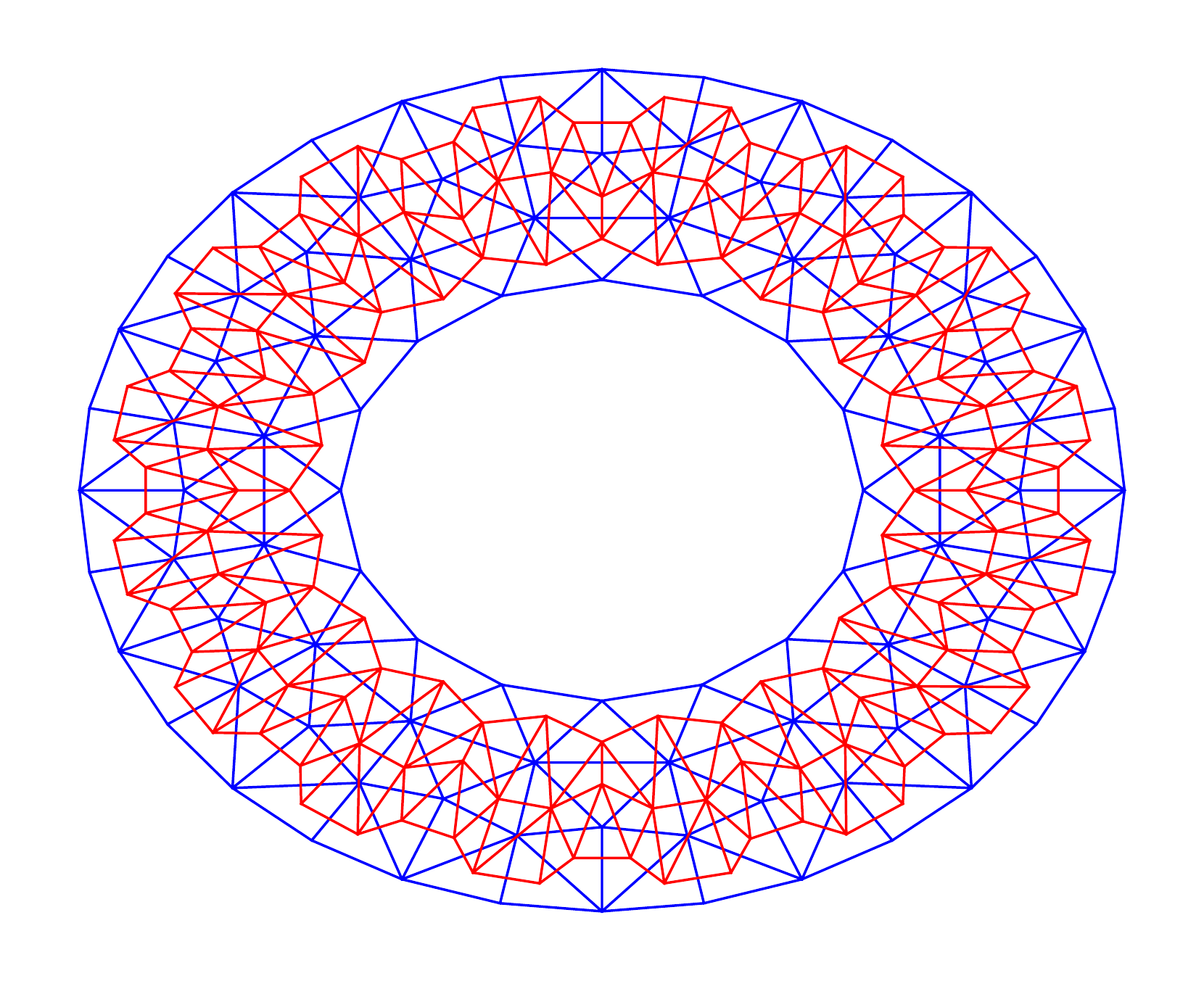}
\caption{Base mesh $\Omega^h$ in blue and dual (lip) mesh $\Delta^h$ in red colour}
\label{fig:3}
\end{figure}

The domain $\Omega$ is discretized using (linear) triangular mesh elements denoted  $\Omega^h$. The displacement field is then discretized using classical finite element discrete space over $\Omega^h$. In that case, the displacement field is continuous over $\Omega^h$ and linear over each element. The strain is then piecewise constant over each element. The internal strains $ \bm{\varepsilon_i}$ are stored at element centroids and piecewise constant over each element.  Following the Galerkin method, the admissible spaces for the discretized displacements and internal strains are 
\begin{align}
    \mathbf{u}^h(t) \in U_m^h(\Omega^h) \subset U_m(\Omega)  \hspace{1cm}
     \bm{\varepsilon}_i^h \in P^h(\Omega^h) \subset P(\Omega) \label{eq:eq55}
\end{align}
where the spaces $U_m^h$ and $P^h$ are the finite-dimensional approximation to the function spaces $U_m$ and $P$ (for more details see \cite{Hughes}).
In the case of phase-field, the damage field is discretized using classical finite element space functions over $\Omega^h$, whereas for the lip-field, the damage field is stored at the element centroid. The admissible space for phase-damage is then given by 
\begin{align}
    d^h(t) \in S^h(\Omega^h) \subset S(\Omega)  \label{eq:eq56}
\end{align}
where $S^h(\Omega^h)$ is a finite-dimensional space defined over $\Omega^h$ using linear basis functions. For the lip-damage, the admissible space is given by
\begin{align}
    d^h(t) \in \mathcal{W}^h(\Delta^h) = \mathcal{L}^h(\Delta^h) \cap \overline{A_m^h}(\Delta^h)  \label{eq:eq57}
\end{align}
where $\Delta^h$ denotes a dual (or lip-) mesh constructed by connecting the centroids of the base mesh $\Omega^h$. Figure \ref{fig:3} displays the mesh $\Omega^h$ (blue) and the corresponding lip-mesh $\Delta^h$ (red) for a circular plate with a hole at the centre. The vertices of lip-mesh are the element centroids of base mesh. The purpose of the lip-mesh is to define the Lipschitz constraints. The lip-damage field is then piecewise constant over $\Omega^h$ and linear over $\Delta^h$. The space $\mathcal{L}^h(\Delta^h)$ and ${\overline{A_m^h}(\Delta^h)}$ are given as follows:
\begin{align}
    &\mathcal{L}^h(\Delta^h) = \{ d \in S^h(\Delta^h) \;:\; \| ( \nabla d)_t\| \leq \frac{1}{l_2},\;\; \forall\; t \in \mathcal{T} (\Delta^h) \} \label{eq:eq58}\\
    &\overline{{A_m^h}}(\Delta^h) =\{ d \in S^h(\Delta^h): \underline{d} \leq d \leq \overline{d} \} \label{eq:eq59}
\end{align}
where $S^h(\Delta^h)$ is the  finite-dimensional space defined over $\Delta^h$ using linear basis functions. $\mathcal{T}(\Delta^h)$ is set of all elements in $\Delta^h$ with $(\nabla d)_t = \mathbf{B}_t \mathbf{d}_e$. $\mathbf{B}_t$ and $\mathbf{d}_e$ are the elemental gradient operator and elemental damage vector defined for an element $t \in \mathcal{T}(\Delta^h)$. Spaces different from $\mathcal{L}^h(\Delta^h)$ had also been considered in \cite{Chevaugeon} to impose the Lipschitz constraint in a discrete setting, but it was reported in the later that the space defined by Eq. (\ref{eq:eq58}) was relatively less prone to mesh orientation and also benefits from the least number of discrete Lipschitz constraints (equal to the number of elements in lip-mesh).
The step 1 to find the local damage field $d_{loc}^h$ over the vertices of the lip-mesh $\Delta^h$ (given by Eq .(\ref{eq:eq48})) is performed using the open source \textit{scipy} \cite{scipy} package of \textit{python}. The bounds estimate associated with $d_{loc}^h$ (given by Eq. (\ref{eq:eq47})) is performed using a \textit{Dijkstra based fast marching algorithm} explained in \cite{Chevaugeon}. This allows to compute the bounds in $O(r\; log( r))$ computations, with $r$ being the number of vertices in Lip-mesh. The estimation of bounds allows us to define the active zone in $\overline{\Delta^h}$ over the lip-mesh as per Eq. (\ref{eq:eq52}). In the inactive zone $\Delta^h/ \overline{\Delta^h}$, the lip-damage field that minimizes the incremental potential is same as $d_{loc}^h$ (property Eq.(\ref{eq:eq51})). The problem that remains to be solved is to find $d^h \in \mathcal{W}^h(\overline{\Delta^h})$ by minimizing the incremental potential over the active zone $\overline{\Delta^h}$. In contrast to the phase-field implementation considered, we follow a direct minimization using the \textit{cp} solver of the open source \textit{cvxopt} \cite{cvxopt} package of \textit{python}. This package allows to impose the discrete constraints $d^h \in \mathcal{W}^h(\overline{\Delta^h)} $ using  the first and second order cone constraints, while minimizing the incremental potential. A direct minimization using an optimisation package in this case is possible (or handy), because the (non-local) minimization is performed only over a small domain $\overline{\Delta^h}$. This is an advantage of the lip-field approach, while if such operation has to be performed for the available phase-field models (to preserve the variational structure) results in quite expensive computational cost, as the minimization has to be performed over the entire domain $\Omega^h$.

\section{Simulation results}
\label{sec:sr}
In this section, numerical results are presented for the bidimensional studies, and we demonstrate the capability of the two different regularization approaches to provide similar results. All the tests are performed using the plane strain assumption with constant Poisson's ratio under constant loading rates. The above assumption allows us to use the following relation: $\lambda_i = E_i \nu /(1+\nu)/(1-2 \nu)$ and $\mu_i = E_i/2/(1+\nu)$ to find the Lame's constants associated with each KV unit. In all the tests, the fracture parameters are set using Eq. (\ref{eq:eq34}) and (\ref{eq:eq35}) to achieve some equivalence between both the lip-field and phase-field approaches. As mentioned earlier, the damage initiation starts at the onset of loading. In order to capture the crack properly, the mesh is refined in the critical zone where the crack is expected to propagate. For the phase-field model considered, it was reported  that the fracture resistance  parameter $G_c$ is overestimated and hence an analogous parameter  to nullify this issue have been used \cite{Bourdin2007} . This replaces $G_c$ in phase-field models (in Eq. (\ref{eq:eq45})) by $G_{c,eff} = G_c/ (1+ \kappa)$. The parameter $\kappa = h/(4l_1)$, where $h$ is the effective element size in the critical zone. The mesh is generated using \textit{gmsh} software \cite{gmsh} and the associated lip-mesh using the python \textit{triangle} package \cite{triangle}. The code to obtain the simulation results can be downloaded from \textit{https\string://github.com/rajasekar2808/ve\string_fracture}.

\subsection{Test A:}
\begin{figure}[H]
  \centering
  {\includegraphics[trim=1cm 14.5cm 1cm 1cm,clip,scale= .5]{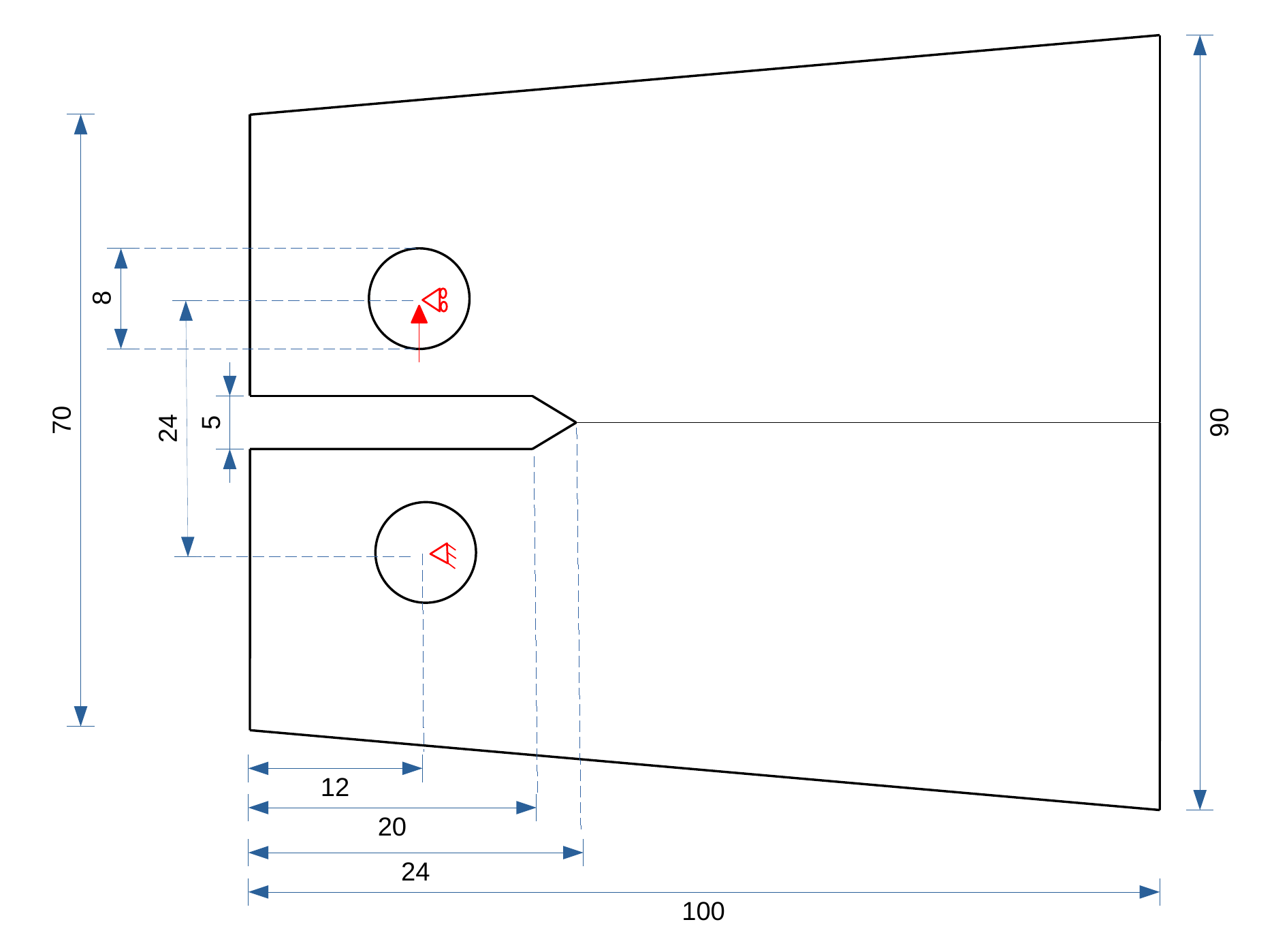}}
  \caption{Test A: Tapered Double Cantilever Beam (TDCB) Geometry specimen} (all dimensions in $mm$) \label{fig:4}
\end{figure}

\begin{figure}[H]
\centering
\captionsetup{justification=centering}
\includegraphics[scale = .5]{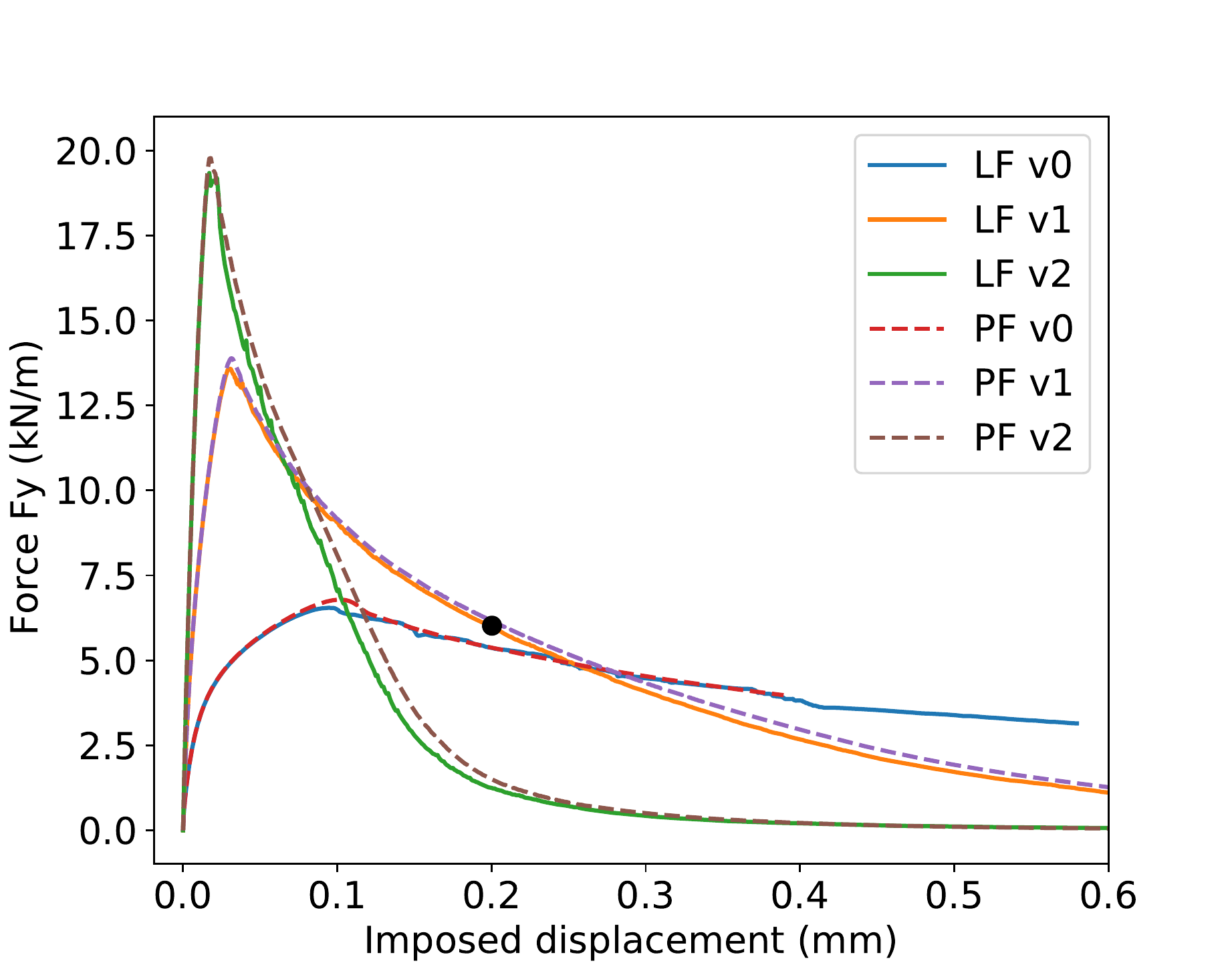}
\caption{Force vs imposed displacement plot for Test A at different loading rates: \\PF and LF indicate Phase and Lip-Field}
\label{fig:5}
\end{figure}

Consider a Tapered Double Cantilever Beam (TDCB) with two symmetrical holes. The geometry and loading conditions are shown in Figure \ref{fig:4}. This geometry is a classical example used to study crack propagation as the tapered beam offers a stable crack growth and the same has been used in \cite{Chevaugeon} to demonstrate the ability of the lip-field approach to simulate Griffith fracture.  We set $\beta=1$ (symmetric tension / compression behavior) in this case.  Here, the critical zone lies in the center strip.  Regarding the loading conditions, the bottom hole has its center fixed and is free to rotate around the z-axis, while the top hole has it center fixed on x-axis and free to rotate around z-axis. The loading is imposed on the top hole along the y-axis through constant loading rates. Three different loading rates are considered for this study $(v0,v1,v2)=(.01,.1,1)$ mm/s. The time steps used for the respective cases are $(.01,.005,.001)$ s. The viscoelastic material is described with 10 KV units (see Table \ref{Table:1} )  with $\nu =.2$ . The parameters associated to fracture are shown in Table  \ref{Table:2}.
\begin{table}[H]
\begin{center}
\scalebox{0.8}{
\begin{tabular}{ | m{1.65cm} | m{.8cm}| m{.8cm} |  m{1cm} | m{.8cm}| m{.8cm} | m{.8cm} | m{.8cm}| m{.8cm} | m{.8cm} |m{.8cm} |m{.8cm} |m{.8cm} |} 
  \hline
 KV unit (i)  & 0 & 1 & 2 & 3 & 4 & 5 & 6 & 7 & 8 & 9 & 10 & 11\\
 \hline
 $E_i$ (MPa) & 31770 & 87398 & 123414 & 65830 & 62457 & 62661 & 7305 & 12500 & 418 & 1743 & 79 & 39 \\
 \hline
 $\tau_{i,{ref}}$ (s) & N/A & 1E-5 & 1E-4 & 1E-3 & 5E-3 & 1E-2 & 1E-1 & 1 & 10 & 1E2 & 5E2 & 1E3 \\
 \hline
 \end{tabular}}
  \caption{Viscoelastic material parameters for Test A}
  \label{Table:1}
\end{center}
\end{table}
\begin{table}[H]
    \begin{center}
    \scalebox{0.8}{
    \begin{tabular}{| m{2cm}| m{2cm} |m{2cm} |m{2cm} |m{2cm} |}
    \hline
         $Y_c$ ($J/m^3$) & $G_c$ ($J/m^2$)  & $l_1$ (mm) & $l_2$ (mm) & $h$ (mm) \\
         \hline
         14e3 & 46.667 & 1.25 & 2.5 & .4 \\
         \hline
    \end{tabular}}
    \caption{Fracture parameters for Test A}
    \label{Table:2}
    \end{center}
\end{table}
Figure \ref{fig:5} shows the load vs imposed displacement for different loading rates, obtained using both lip-field and phase-field regularization. It is evident that both of these regularization techniques produce similar results. This is indeed expected as a result of some equivalence derived in section \ref{sec:eq}. 
It can be observed that for all the loading rates considered, the respective reaction forces initially increases until it reaches a critical displacement, after which the reaction force drops in a controlled manner. The viscoelastic description of the material also induces some rate effects, with the stiffening of material  and simulating elastic behaviour at higher displacement rates. Moreover the material ruptures relatively faster for higher imposed displacement rates, because of the material stiffening close to crack and the available free energy being spent relatively more for damage dissipation compared to viscous dissipation. The incremental energy balance between two consecutive time steps is also demonstrated in Appendix \ref{ap:d}.

\begin{figure}[H]
\centering
\captionsetup{justification=centering}
\includegraphics[scale = .3]{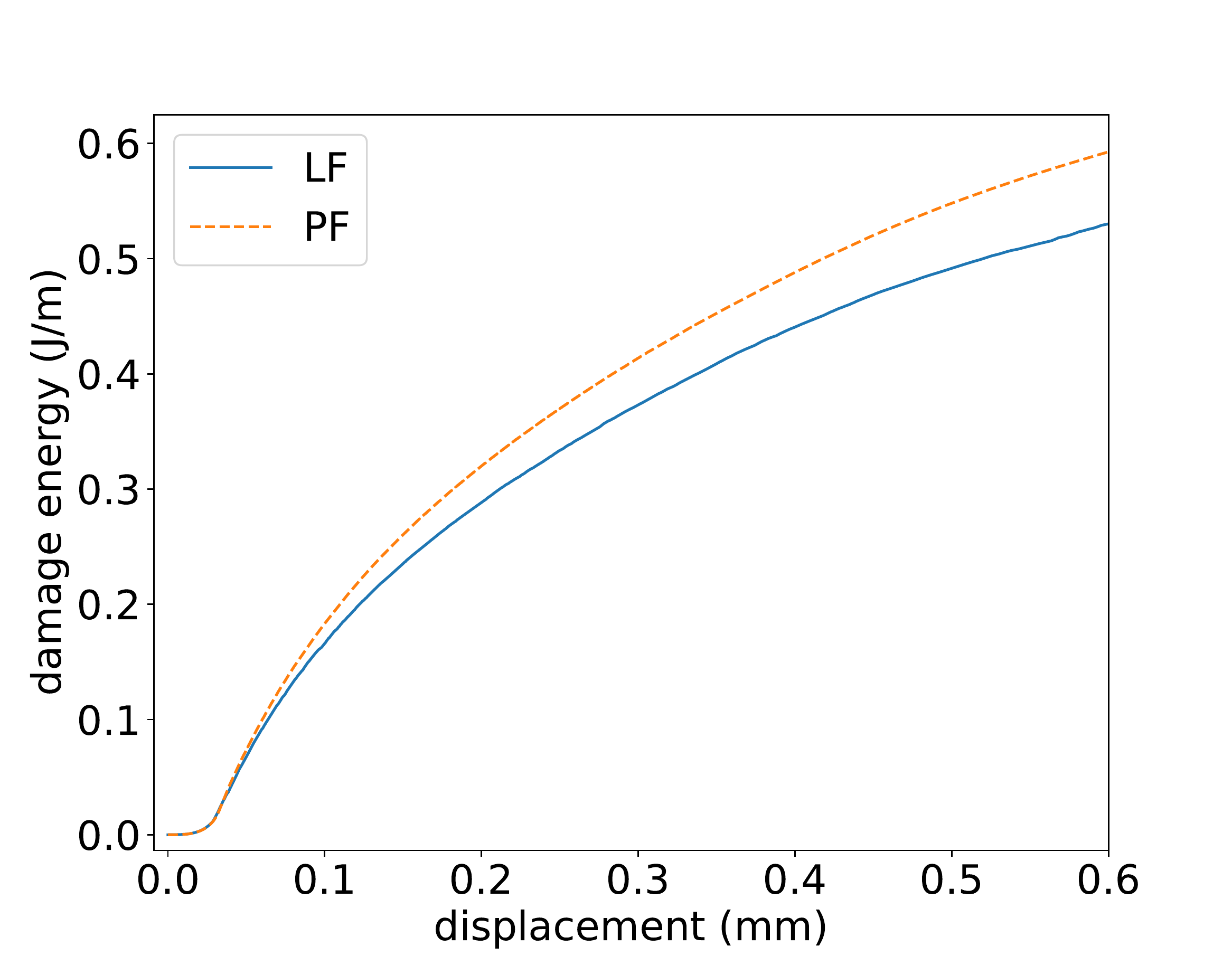}
\caption{Phase-field (PF) and lip-field (LF) damage energies of the bulk as a function of the imposed displacement for test A with the imposed displacement rate $v1$.}
\label{fig:7}
\end{figure}

\begin{figure}[H]
  \centering
  {\includegraphics[trim=1cm 21.5cm 1cm 1.5cm,clip,scale= .8]{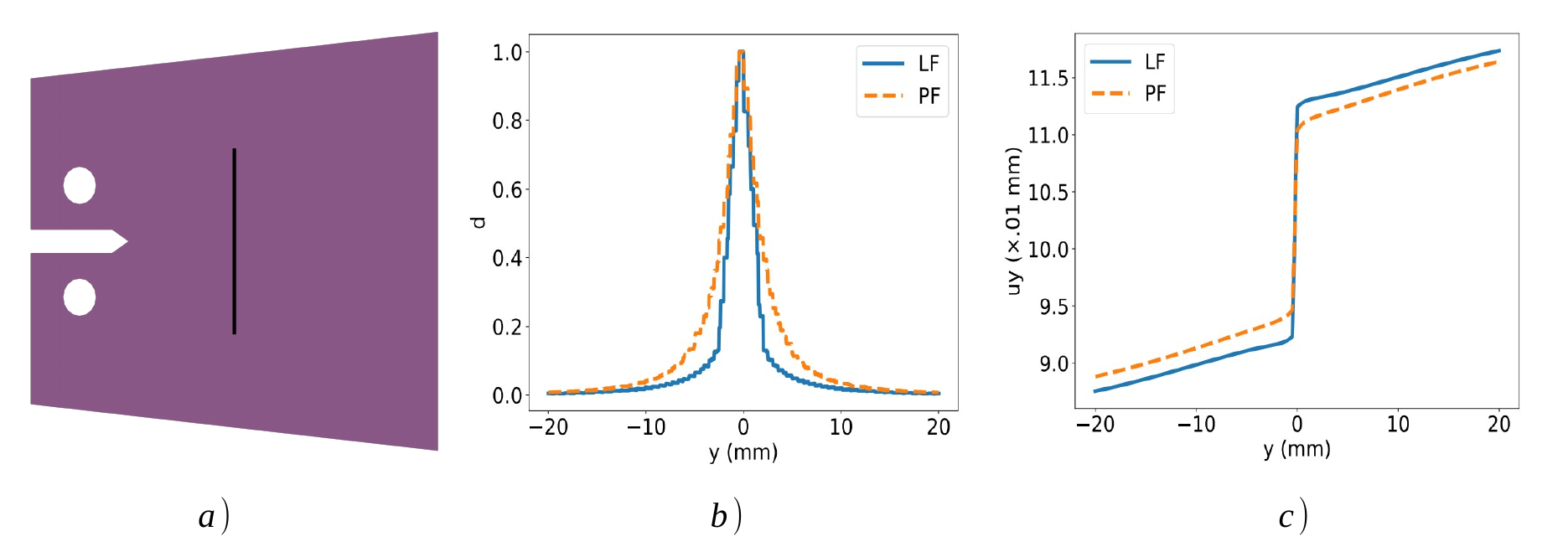}}
  \caption{ Damage profiles (b) and displacement in y-direction (c)  along the black section in (a)}\label{fig:8}
\end{figure}

Figure \ref{fig:6} plots the contour profiles for both damage and stress in the y-direction for the imposed displacement rate $v1$ at the anchor point indicated in Figure \ref{fig:5}. The stress plot is superimposed over the deformed mesh with the elements corresponding to $d>.99$ being removed.
\begin{figure}[H]
  \centering
  \captionsetup{justification=centering}
  {\includegraphics[trim=1cm 12cm 1cm 2cm,clip,scale= .6]{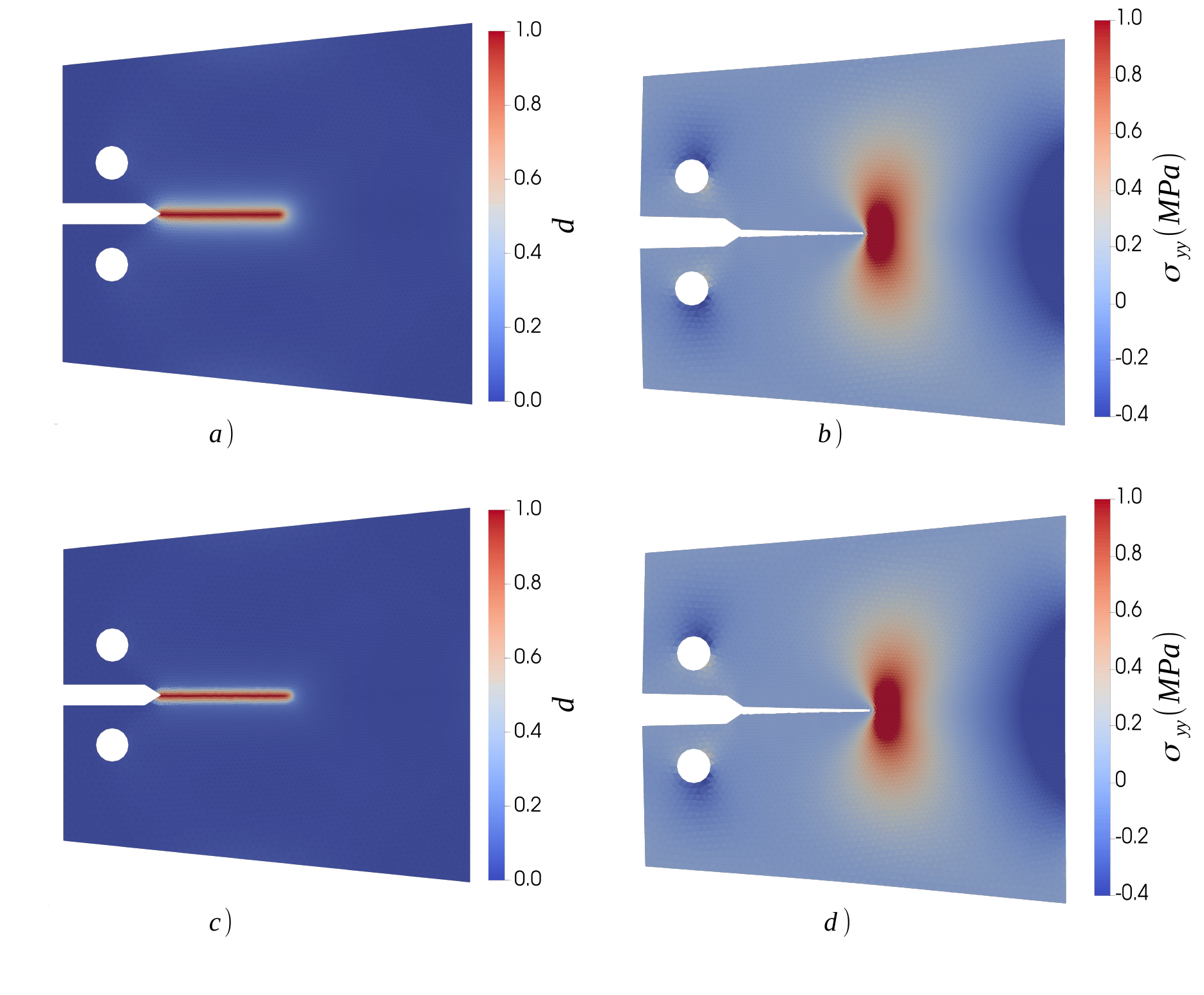}}
  \caption{Damage (left) and stress-y (right) contours for phase-field (top row) and Lip-field (bottom row). Contour plotted for Test A with imposed displacement rate $v1$ at the anchor point indicated in Figure \ref{fig:5}. The stress is plotted over the deformed mesh with magnification factor = 2. }\label{fig:6}
\end{figure}
The crack can be observed in the center strip as a band of localized elements with damage extending to 1. Figure \ref{fig:7} plots the damage energy in bulk for both phase-field and lip-field as a function of imposed displacement. The damage energy is calculated using the relation $\int_{\Omega} \Tilde{\phi}_d\;d\Omega$.  It can be seen that the damage dissipation is higher in the case of PF for a given imposed displacement. However, the crack length in the case of PF is relatively little shorter (from Figure \ref{fig:6}) compared to LF, despite the fact that PF is more dissipative (see Figure \ref{fig:7}). This can be explained by the exponential damage profile resulting in damage diffusion happening perpendicular to the crack path for PF. 
For the anchor point shown in Figure \ref{fig:5}, the section profiles for damage and displacement in y-direction is plotted on Figure \ref{fig:8}.
PF model results in some damage widening due to the diffusive nature as explained before. In case of LF, the displacement jump is also observed to be relatively more across the band of localized elements because of the higher slope of the lip-damage field across the crack.
It can also be seen that the lip-damage profile is not exactly linear in this case. This is due to the damage initiation at the onset of loading, which causes the lip-damage to spread initially before localizing. However, the spread is minimal relative to the exponential damage profile of PF.

\subsection{Test B}
\begin{figure}[H]
  \centering
  {\includegraphics[trim=1cm 20.5cm 1cm 1.5cm,clip,scale= .7]{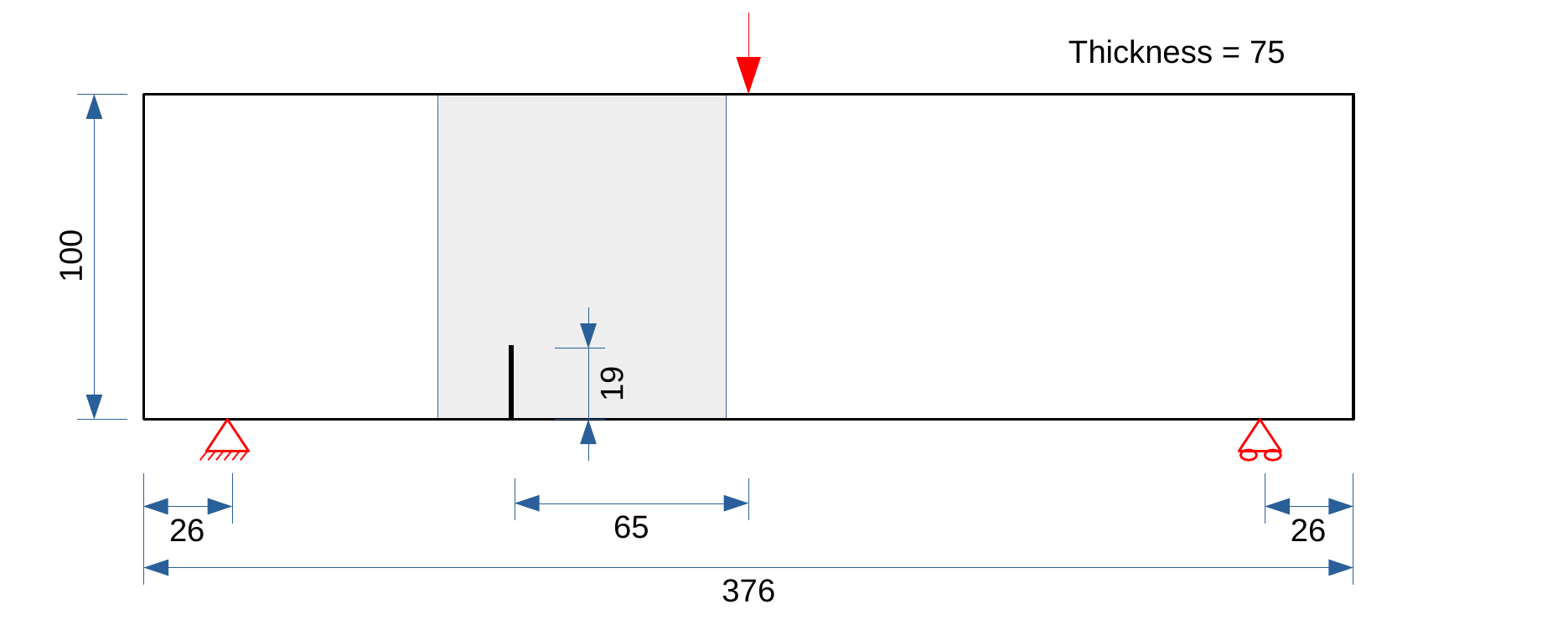}}
  \caption{Test B: 3 point bending specimen with notch located at offset (all dimensions in $mm$) }\label{fig:9}
\end{figure}

\begin{figure}[H]
  \centering
  {\includegraphics[trim=4cm 10cm 5cm 11cm,clip,scale= .7]{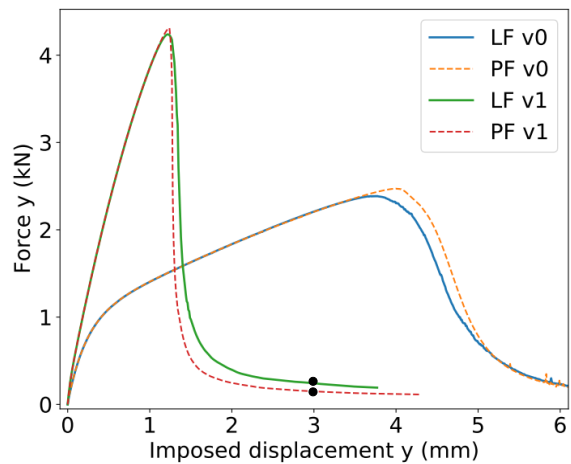}}
  \caption{Force vs imposed displacement plot for Test B at different loading rates: \\PF and LF indicate Phase and Lip-Field  }\label{fig:10}
\end{figure}

\begin{figure}[H]
  \centering
  \captionsetup{justification=centering}
  {\includegraphics[trim=1cm 12cm 1.5cm 1.5cm,clip,scale= .6]{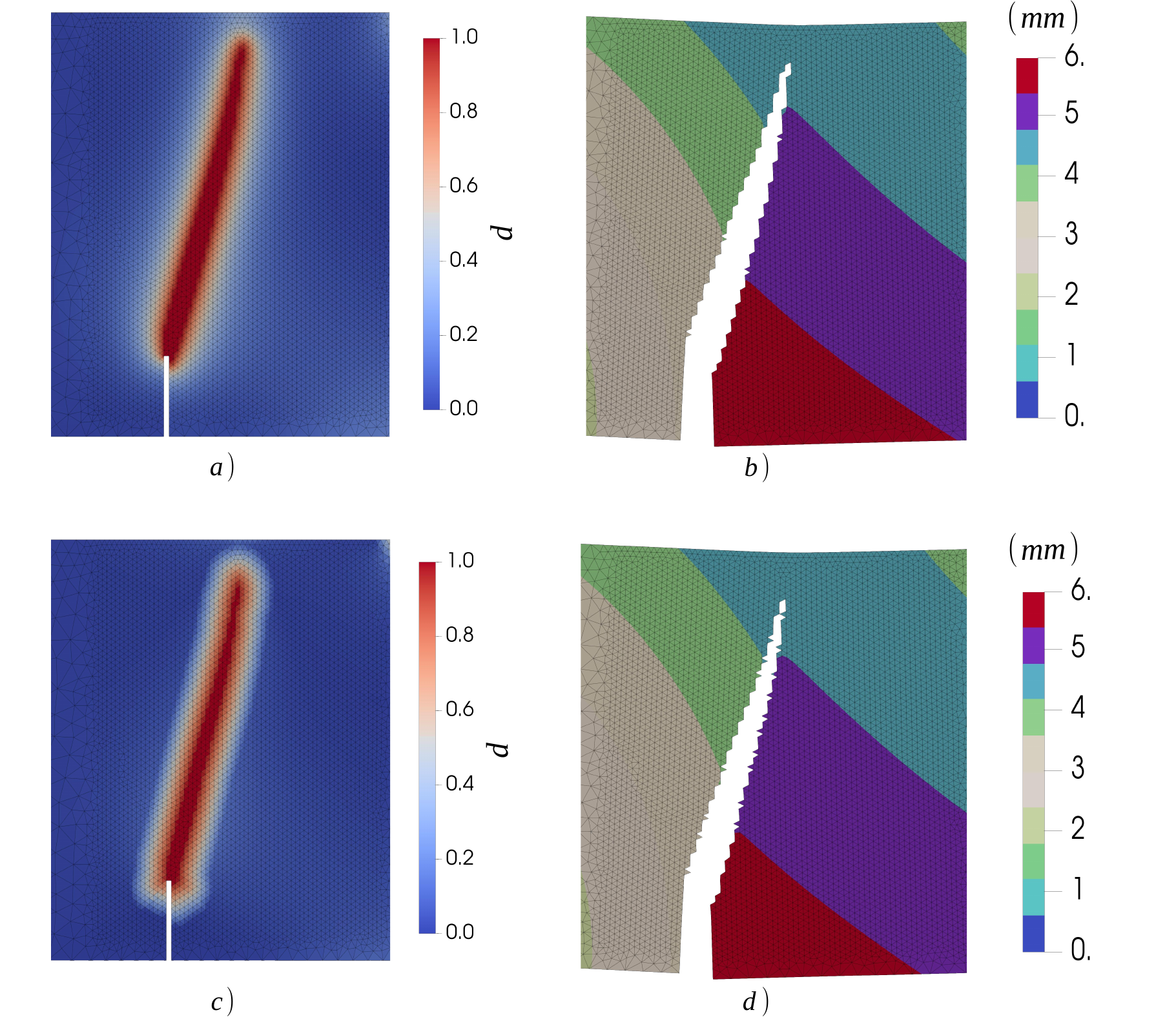}}
  \caption{Damage (left column) and magnitudes of displacement (right column) contours for phase-field (top row) and lip-field (bottom row). Contour plotted for Test B with imposed displacement rate $v1$ at the anchor point indicated in Figure \ref{fig:10}. The magnitudes of displacement is plotted over the deformed mesh with magnification factor= 2.}\label{fig:11}
\end{figure}

This test is made up of a three-point bending beam with a notch located at offset. The geometry and loading conditions are represented in the Figure \ref{fig:9}. This is considered as one of the classical test for mixed mode fracture in case of viscoelasticity \cite{Buttlar}. Since the beam is subjected to both tension and compression, $\beta =0$ is used to prevent damage growth in compression. The critical zone is denoted in Figure \ref{fig:9} using a shaded area. To avoid damage concentration at the  boundary  conditions, boundary conditions are applied over a strip of small thickness. Two different loading rates are considered for this study $(v0,v1) = (.1,1)$ mm/s with the respective time step $(.1,.01)$ s.   The viscoelastic material is described with 3 KV units (see Table \ref{Table:3})  with $\nu =.2$ . The parameters associated with the fracture are shown in Table \ref{Table:4} 
\begin{table}[H]
\begin{center}
\scalebox{0.8}{
\begin{tabular}{ | m{1.65cm} | m{.8cm}| m{.8cm} |  m{1cm} | m{.8cm}|} 
  \hline
 KV unit (i)  & 0 & 1 & 2 & 3 \\
 \hline
 $E_i$ (MPa) & 2300 & 1500 & 800 & 100  \\
 \hline
 $\tau_{i,{ref}}$ (s) & N/A & .05 & 15 & 26\\
 \hline
 \end{tabular}}
  \caption{Viscoelastic material parameters for Test B and C}
  \label{Table:3}
\end{center}
\end{table}
\begin{table}[H]
    \begin{center}
    \scalebox{0.8}{
    \begin{tabular}{| m{2cm}| m{2cm} |m{2cm} |m{2cm} |m{2cm} |}
    \hline
         $Y_c$ ($J/m^3$) & $G_c$ ($J/m^2$)  & $l_1$ (mm) & $l_2$ (mm) & $h$ (mm) \\
         \hline
         14e3 & 186.667 & 5 & 10 & 1.4 \\
         \hline
    \end{tabular}}
    \caption{Fracture parameters for Test B}
    \label{Table:4}
    \end{center}
\end{table}
The reaction force at supports vs imposed displacement is plotted on Figure \ref{fig:10}. Similar to the observation made in Test A, both the regularization techniques produce results in good agreement with each other. We can also observe the rate effects, with the crack propagation being relatively slow in case of lower imposed displacement rates and brutal character of crack propagation at higher displacement rates. 
Figure \ref{fig:11} plots the damage and magnitude of displacement (over the deformed mesh) corresponding to the imposed displacement rate $v1$ at the anchor point indicated in Figure \ref{fig:10}. The band of elements with localized damage values can be considered as an approximation of  the macro-crack. Hence, the crack is represented in Figure \ref{fig:11}.b and \ref{fig:11}.d by removing elements corresponding to $d>.99$. The crack profile observed is similar to the results observed in \cite{Buttlar} for the asphalt material. The damage profile for the phase-field is observed to be more diffusive because of the exponential damage profile. Concerning the displacement profile, a displacement jump is clearly visible for both the approaches. However, small discrepancies in the jump along the crack path is observed due to discrepancies in damage profile.

\subsection{Test C}
\begin{figure}[H]
  \centering
  \captionsetup{justification=centering}
  {\includegraphics[trim=1cm 1.5cm 1.5cm 20cm,clip,scale= .8]{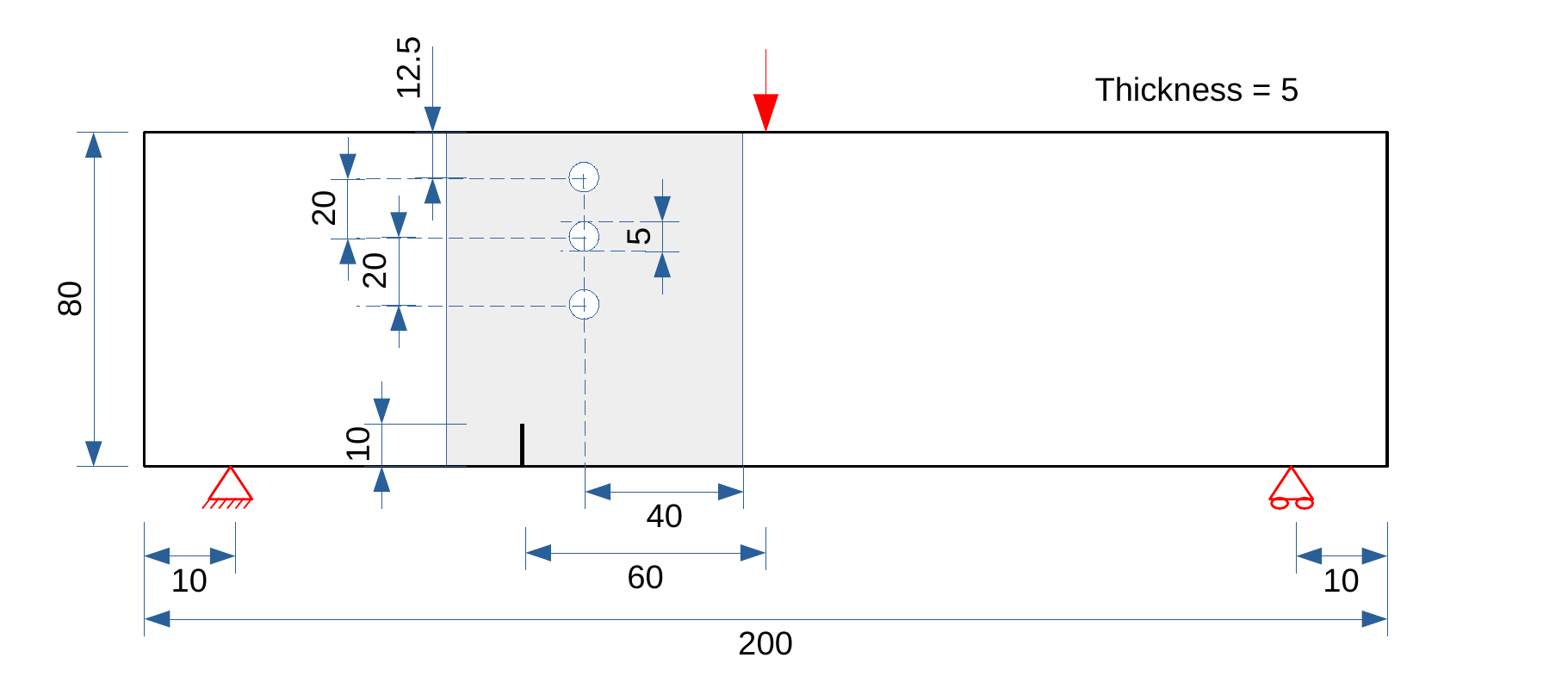}}
  \caption{Test C: 3 point bending beam specimen with holes and notch located at offset (all dimensions in $mm$) }\label{fig:12}
\end{figure}

\begin{figure}[H]
  \centering
  {\includegraphics[trim=1cm 11cm 1.5cm 1.5cm,clip,scale= .7]{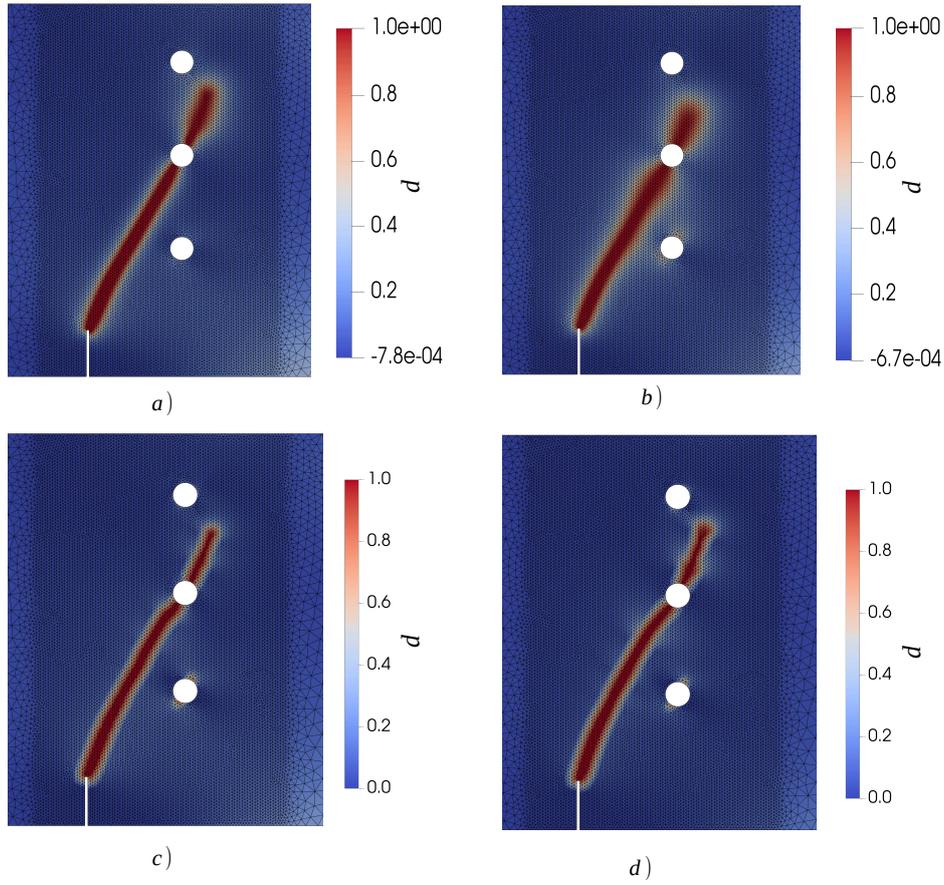}}
  \caption{Phase damage (top row) and lip damage (bottom row) contours for imposed displacement rates $v0$ (left column) and $v1$ (right column) close to rupture. }\label{fig:13}
\end{figure}

This test is similar to the test B with a notch located at offset, except for the addition of holes along the crack path. The geometry and loading conditions are sown in Figure \ref{fig:12}. $\beta =0$ is again used to prevent damage growth in compression. The same test has been considered in \cite{Miehe2} and \cite{Cazes}. In the later, this test has been used to compare the Thick Level Set (TLS) approach and a particular phase-field approach (AT2 model). The parameters for the viscoelastic material remain the same as in Test B (Table \ref{Table:3}). The parameters associated with the fracture are listed in Table \ref{Table:5}. Two different imposed displacement rates $(v0,v1)=(.05,1)$ mm/s with respective time steps $(.5,.005)$ s are considered. The damage is imposed zero close to supports and loading points to avoid damage growth due to concentrated loading. 
\begin{table}[H]
    \begin{center}
    \scalebox{0.8}{
    \begin{tabular}{| m{2cm}| m{2cm} |m{2cm} |m{2cm} |m{2cm} |}
    \hline
         $Y_c$ ($J/m^3$) & $G_c$ ($J/m^2$)  & $l_1$ (mm) & $l_2$ (mm) & $h$ (mm) \\
         \hline
         14e3 & 50.4 & 1.35 & 2.7 & .3 \\
         \hline
    \end{tabular}}
    \caption{Fracture parameters for Test C}
    \label{Table:5}
    \end{center}
\end{table}
In this case, we are only interested in the damage profiles for the phase-field and lip-field approaches at the considered loading rates. The damage contour is plotted in Figure \ref{fig:13}. The crack profile observed is similar to the profiles observed in \cite{Miehe2},\cite{Cazes} for quasi-brittle fracture. 
It is observed that the loading rate did not have any significant effect on the crack orientation. However, an increase in the width of the damage band with an increase in loading rate is observed for the PF model despite the same regularizing length scales being used for both loading rates. This might be due to the brutal character of crack propagation in case of increased loading rate, while for lower loading rate the crack propagation is less brutal because of the viscous dissipation acting as a resistance to damage growth (as a result of energy conservation in bulk).  A similar effect was also observed in \cite{Miehe2}  for the case of phase-field fracture in elasticity when some artificial viscous resistance was added to $G_c$.  Regarding the boundary effects as seen in Figure \ref{fig:13}, gradient of damage is not necessary zero along the normal to the boundary in case of LF, whereas in case of PF, it is observed that gradient of damage is zero along the normal to the boundary because of the Neumann condition as a result of Eq. (\ref{eq:eq44}). This also results in a swelling of phase damage close to the mid hole for the considered mesh. It is expected to be overcome by reducing the length scale. Besides, small negative values of damage were observed in case of phase-field models (see Figures \ref{fig:13} a,b), showing that the irreversibility constraint is not properly respected. This was also observed in the case of Tests A and B for the phase-field models (but not reported as the negative values were close to zero). This is because of the use of the local history field to drive fracture, which results in replacing variational inequality by a variational equality. As a consequence, $\dot{d} \geq 0$ is not properly respected. However, the discrepancy is small that it is ignored in most studies. While the exact variational structure of the problem is being preserved in lip-field case, the damage irreversibility constraint is properly enforced and damage lies strictly between 0 and 1.
\section{Conclusion}
\label{sec:cn}
We presented a variationally consistent formulation to viscoelastic fracture in the quasi-static case by use of damage models. Both Phase-Field (PF) and Lip-Field (LF) regularization approaches have been considered to alleviate the spurious localization problem of softening damage models. In addition, the LF approach has been used for the first time in a viscoelastic context. 

Despite both these approaches share common aspects that they are variational and posses similar expression for energy in local form, they are fundamentally different : (i) In PF, the energy depends on the gradient of damage to introduce regularization, while in case of LF, the local expression of energy is maintained. The regularization for the later is then through the introduction of a Lipschitz continuous space (non-local) for the lip-damage field. (ii) Phase-damage in the AT2 type models is sought by solving a partial differential equation with its questionable boundary condition that gradient of phase-damage is zero on the boundaries. However, in LF approach, a direct optimization of the energy to find lip-damage is practical, thanks to the local-non local split of the minimization process \cite{Moes LF}. This allows to perform the non local minimization only over a small portion of the domain affected by Lipschitz constraint.

Compared to the PF implementation, the variational structure of the problem is preserved in case of LF, as the objective function used for minimization is not modified. In contrast, for the PF implementation, the use of a local history variable to impose damage irreversibility constraint results in changing the objective function and hence results in loss of the variational structure. The negative values of damage observed for the numerical simulation also indicates that the history variable doesn't impose irreversibility constraint in a strong sense.

In order to aid the comparison of results of both the PF and LF approaches for viscoelastic fracture, some equivalence is sought by having similar damage profile and fracture energies. Numerical simulations indicate both these approaches compare well. However in case of PF models, damage widening is observed. This is due to the diffusive nature of the PF model considered. In addition, for the PF model in study, the questionable boundary condition that gradient of damage should be zero on boundaries has been observed to have swelling effect of damage close to boundaries in some cases. 

It has also been observed that both the models were able to capture the rate effects due to viscoelastic nature of bulk while preserving energy conservation. In our case, the choice of damage dissipation potential limits us to a particular type of PF and LF models. However, having some equivalence derived for any other given PF and LF models, it is expected that the similarities/differences pointed out in this work still applies. Finally, the use of such viscoelastic damage models for practical applications like fracture in asphalt and the extension to dynamic case would also be an interesting topic. This will be the objective of the future study.

\appendix
\section{Appendix A}\label{ap:a}
In this section the expression for eigen split of the strain tensor $ \bm{\varepsilon}$ is listed
\begin{align}
    \bm{\varepsilon} =  <\bm{\varepsilon}>_+ \; + \;< \bm{\varepsilon}>_{-}  \label{eq:eq60}
\end{align}
where $< \bm{\varepsilon}>_+$ and $< \bm{\varepsilon}>_{-}$ describe the tensile and compressive modes. They are defined based on the eigen values ($\{ \varepsilon^a \}_{a=1,2,..,\beta}$) and eigen vectors ($\{ \mathbf{n}^a \}_{a=1,2,..,\beta}$) of $ \bm{\varepsilon}$. $\beta$ takes the values 2 and 3 for two and three dimensions respectively. Then, the expression for the  tensile and compressive modes is given as follows:
\begin{align}
    < \bm{\varepsilon}>_+ = \sum_{a=1}^{\beta} <\varepsilon^a>_+ \mathbf{n}^a \otimes \mathbf{n}^a \hspace{1cm} < \bm{\varepsilon}>_{-} = \sum_{a=1}^{\beta} <\varepsilon^a>_{-} \mathbf{n}^a \otimes \mathbf{n}^a \label{eq:eq61}
\end{align}
with the following definitions of bracket operators:
\begin{align}
    <x>_+ = \frac{1}{2}( x + |x|) \hspace{1cm} <x>_{-} = \frac{1}{2}( x - |x|)  \label{eq:eq62}
\end{align}

\section{Appendix B}
\subsection{Appendix B.1} \label{ap:b1}
For the symmetrical case ($\beta=1$), the free energy and dissipation potential are  given by Eq. (\ref{eq:eq23}) and Eq. (\ref{eq:eq21}). They can be rewritten as follows:
\begin{align}
    &\psi =  g(d) \frac{1}{2} \left( \bm{\varepsilon} - \sum_{i=1}^n \bm{\varepsilon_i} \right): \mathbb{C}_0 : \left( \bm{\varepsilon} - \sum_{i=1}^n \bm{\varepsilon_i} \right) \;+\; \sum_{i=1}^n g(d)  \frac{1}{2}\bm{\varepsilon_i}:\mathbb{C}_i : \bm{\varepsilon_i}  \label{eq:eq64}\\
    &\phi = \sum_{i=1}^n \frac{1}{2} \bm{\dot{\varepsilon_i}}:\tau_i \mathbb{C}_i : \bm{\dot{\varepsilon_i}} \label{eq:eq65}
\end{align}
where $\mathbb{C}_i$ is the fourth order elastic tensor associated to $i^{th}$ KV unit (Figure \ref{fig:1}) and is given by $\mathbb{C}_i = 2 \mu_i \mathbb{I} + \lambda_i \mathbf{1} \otimes \mathbf{1}$. ($\mathbb{I}$ and $\mathbf{1}$ are the fourth and second order identity tensors respectively.)

At time step $t_{m+1}=t$, the stress ${ \bm{\sigma}}_{m+1}= { \bm{\sigma}}$ is given by
\begin{align}
    { \bm{\sigma}} = \frac{\partial \psi}{\partial { \bm{\varepsilon}}} = g(d) \mathbb{C}_{0} : \left( { \bm{\varepsilon }} - \sum_{i=1}^n { \bm{\varepsilon_i }}  \right) \label{eq:eq66}
\end{align}
The stress in $i^{th}$ KV unit is given by the sum of the stresses in its elastic ($\bm{\sigma_{e,i}}$) and viscous part ($\bm{\sigma_{\eta,i}}$)
\begin{align}
    { \bm{\sigma}}_i = \bm{\sigma_{e,i}} +  { \bm{\sigma_{\eta,i}}} = g(d) \mathbb{C}_\mathbf{i}  :\bm{\varepsilon_i} + \tau_i \mathbb{C}_\mathbf{i}  :\bm{\dot{\varepsilon}_i}\; \approx \;g(d) \mathbb{C}_\mathbf{i} : \bm{\varepsilon_i} + \tau_i \mathbb{C}_\mathbf{i} :\left(\frac{ \bm{{\varepsilon}_{i}}- \bm{{\varepsilon}_{i,m}}}{\Delta t} \right)  \label{eq:eq67}
\end{align}
Since the KV units are connected in series, the stress in each KV units are the same ($ \bm{\sigma}= \bm{\sigma_i}$). Equating Eq.\ref{eq:eq66} and \ref{eq:eq67} gives the following expression for $ \bm{\varepsilon_i}$ (=$ \bm{{\varepsilon}_{i,m+1}} $)
\begin{align}
     \bm{\varepsilon_i} = \frac{\Delta t}{g(d) \Delta t + \tau_i} \mathbb{C}_i^{-1} : \left(  \bm{\sigma} + \frac{\tau_i}{\Delta t} \mathbb{C}_i : \bm{{\varepsilon}_{i,m}} \right) \label{eq:eq68}
\end{align}
Substituting Eq. (\ref{eq:eq68}) in Eq. (\ref{eq:eq66}) allows us to rewrite $ \bm{\sigma}$ as
\begin{align}
    { \bm{\sigma}} = \mathbb{H} : \left( { \bm{\varepsilon}} - \sum_{i=1}^{n} \frac{\tau_i}{g(d) \Delta t + \tau_i}  \bm{\varepsilon_{i,m}} \right) = \mathbb{H} : \bm{\varepsilon} -  \bm{\sigma_{int}}  \label{eq:eq69}
\end{align}   
where $\mathbb{H}$ and $\bm{\sigma_{int}}$ are given as follows
\begin{align}
    &\mathbb{H} = g(d) \left[ \mathbb{I} + \sum_{i=1}^n \frac{g(d) \Delta t}{g(d) \Delta t + \tau_i} \mathbb{C}_0 : \mathbb{C}_i^{-1} \right]^{-1} : \mathbb{C}_0    \label{eq:eq70}\\
    & \bm{\sigma_{int}} = \mathbb{H}:\left( \sum_{i=1}^{n} \frac{\tau_i}{g(d) \Delta t + \tau_i}  \bm{\varepsilon_{i,m}} \right)\label{eq:eq71}
\end{align}  
Substitution of Eq.(\ref{eq:eq69}) in Eq. (\ref{eq:eq3}) results in the variational form given by Eq. (\ref{eq:eq36}) for finding displacements.

\subsection{Appendix B.2} \label{ap:b2}
For the asymmetric tension/compression behavior ($\beta=0$), the $j^{th}$ iteration of the Newton method is given by
\begin{align}
    \mathbf{K}^j \;  \bm{\Delta\alpha}^j = \mathbf{r}^j  \label{eq:eq72}
\end{align}
where $\mathbf{K}^j$, $\mathbf{r}^j$ and $ \bm{\Delta\alpha}^j$ are the Hessian matrix, the vector of residuals and the vector of unknowns at iteration $j$. $ \bm{\alpha}$ contains the vectors of unknown fields of state variables given by $[\mathbf{u},  \bm{\varepsilon_1},.., \bm{\varepsilon_n}]^T$ and $ \bm{\Delta\alpha}^j =  \bm{\alpha}^{j+1}-  \bm{\alpha}^j$. Rewriting Eq. (\ref{eq:eq72}) gives
\begin{equation} \label{eq:eq73}
\renewcommand{\arraystretch}{1.0}
\left(
\begin{array}{cccc}
  \frac{\partial^2 F^j}{\partial \mathbf{u}^2} & \frac{\partial^2 F^j}{\partial  \bm{\varepsilon_1} \partial \mathbf{u}} & \cdots & \frac{\partial^2 F^j}{\partial  \bm{\varepsilon_n} \partial \mathbf{u} } \\[5pt]
  \frac{\partial^2 F^j}{\partial \mathbf{u}  \partial  \bm{\varepsilon_1} } & \frac{\partial^2 F^j}{\partial  \bm{\varepsilon_1}^2 }  & \cdots & \frac{\partial^2 F^j}{\partial  \bm{\varepsilon_n} \partial  \bm{\varepsilon_1} } \\
  \vdots          & \vdots & \ddots          & \vdots           \\
  \frac{\partial^2 F^j}{\partial \mathbf{u} \partial  \bm{\varepsilon_n} } & \frac{\partial^2 F^j}{\partial  \bm{\varepsilon_1} \partial  \bm{\varepsilon_n} } & \cdots & \frac{\partial^2 F^j}{\partial  \bm{\varepsilon_n}^2 }
\end{array}
\right)
\left\{
\begin{array}{c}
   \bm{\Delta} \mathbf{u}^j \\[5pt]
   \bm{\Delta}  \bm{\varepsilon_{1}^j} \\
  \vdots       \\
   \bm{\Delta}  \bm{\varepsilon_{n}^j}
\end{array}
\right\}
=
\left\{
\begin{array}{c}
  -\frac{\partial F^j}{\partial \mathbf{u}} \\[5pt]
  -\frac{\partial F^j}{\partial  \bm{\varepsilon_1}} \\
  \vdots       \\
  -\frac{\partial F^j}{\partial  \bm{\varepsilon_n}}
\end{array}
\right\}
\end{equation}
where the Hessian matrix $\mathbf{K}^j$ and the residuals $\mathbf{r}^j$ are evaluated using numerical differentiation with the known variables $\mathbf{u}^j,  \bm{\varepsilon_i}^j$. The variables $\mathbf{u}^{j+1},  \bm{\varepsilon_i}^{j+1}$ are found using the following relation.
\begin{align}
    \Delta \mathbf{u}^j = \mathbf{u}^{j+1} - \mathbf{u}^j  \hspace{2cm} \Delta  \bm{\varepsilon_i}^j = \bm{\varepsilon_i}^{j+1} - \bm{\varepsilon_i}^j   \label{eq:eq74}
\end{align}

\section{Appendix C} \label{ap:c}
\begin{figure}[H]
  \centering
  \captionsetup{justification=centering}
  {\includegraphics[trim=1cm 16cm 1cm 2cm,clip,scale= .7]{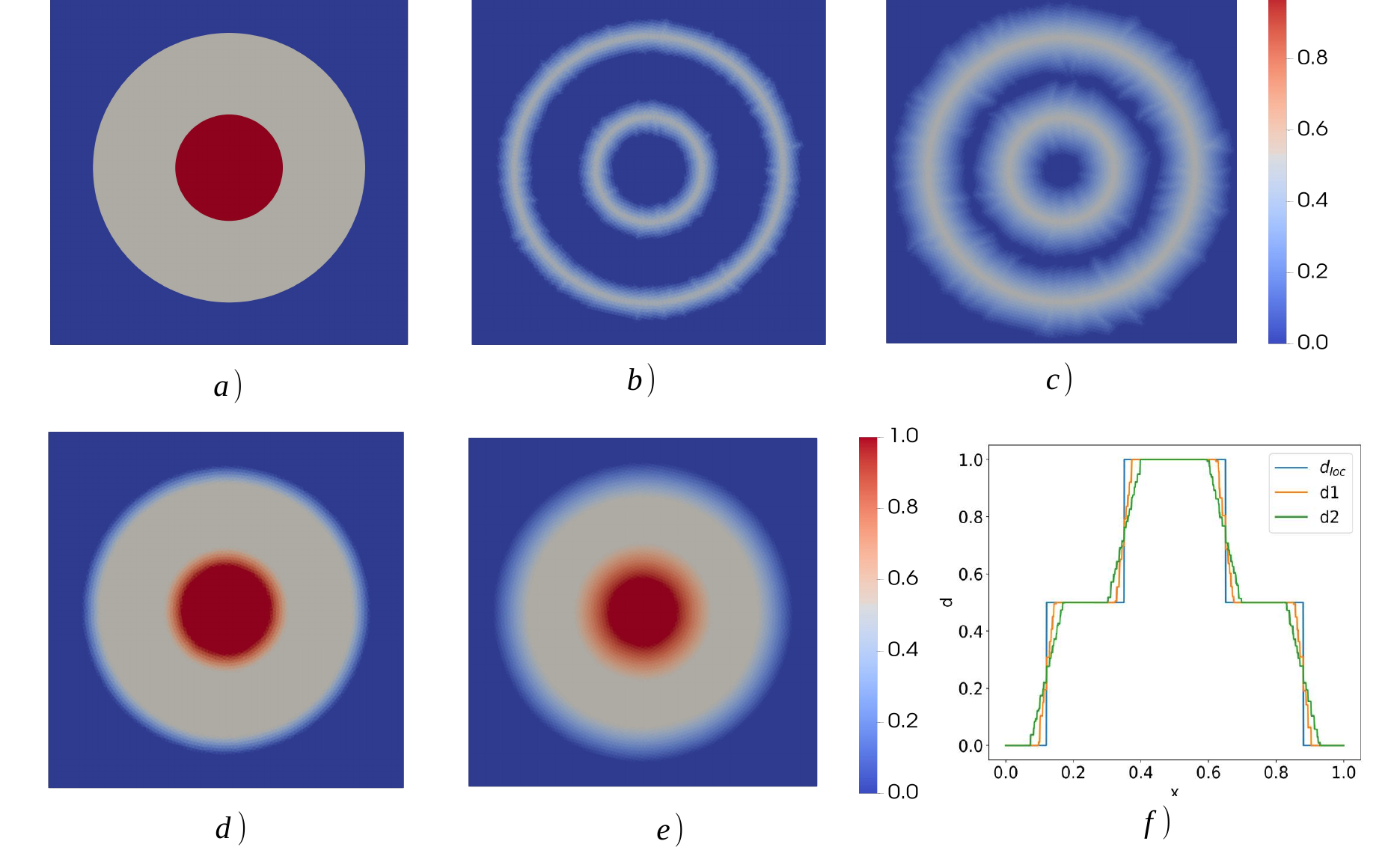}}
  \caption{(a) $d_{tar}$ (or local damage field) with discontinuities ; (b) difference in bounds $\overline{d}-\underline{d}$  for $l_2 = .1$ ; (c)  $\overline{d}-\underline{d}$  for $l_2 = .2$ ; (d) and (e) Lipschitz projection of $d_{tar}$ for  $l_2 = .1$ and $l_2 = .2$ indicated as $d1$ and $d2$; (f) $d_{tar},d1$ and $d2 $ plotted along the mid section }\label{fig:14}
\end{figure}

In this section, we demonstrate the use of bounds to simplify the projection of a given damage field into a Lipschitz space. Consider the problem of projection of a given target (or local) damage field $d_{tar}$ defined over a domain $\Omega$ into a Lipschitz space $\mathcal{L}_{\Omega}$ (given by Eq. (\ref{eq:eq29}) with the regularizing length $l_2$). The aforementioned projection can be reformulated as the following minimisation problem in the sense of $L^2$ norm:
\begin{align}
    d = \arg \min_{d' \in \mathcal{L}_\Omega} \int_{\Omega} |d' - d_{tar}|^2 \;d\Omega   \label{eq:eq75}
\end{align}
Finding the bounds of $d_{tar}$ using Eq. (\ref{eq:eq47}) allows us to define the inactive ($\Omega / \overline{\Omega}$) and active ($\overline{\Omega}$) zones as explained in Section \ref{sec:lf}. The above problem given by Eq. (\ref{eq:eq75}) can then be simplified to the following problem with the  minimisation performed over only on a small part $\overline{\Omega}$ of the original domain $\Omega$ as follows:
\begin{align}
    d = \left\{
                \begin{array}{ll}
                  d_{tar} \hspace{.3cm} &\forall\; \mathbf{x} \in \Omega / \overline{\Omega} \\[1.5pt]
                  \arg \displaystyle \min_{d' \in \mathcal{L}_{\overline{\Omega}}} \int_{\overline{\Omega}} |d' - d_{tar}|^2 \;d\overline{\Omega} \hspace{.3cm} &\forall\; \mathbf{x} \in  \overline{\Omega}
                \end{array}
              \right\}  \label{eq:eq76}
\end{align}
Figure \ref{fig:14} composing a square domain with unit length explains the Lipschitz projection of $d_{tar}$  for two different regularizing length scales $l_2 = (.1,.2)\; units$.  $d_{tar}$ is plotted in Figure \ref{fig:14}.a and it is discontinuous across the two circles of increasing radius. $d_{tar}$ clearly doesn't belong to $\mathcal{L}_\Omega$ because of the discontinuity.
The Figure \ref{fig:14}.b and \ref{fig:14}.c shows the difference in the upper and lower bounds ($\overline{d}-\underline{d}$) associated with $d_{tar}$ for the two different regularizing lengths considered. The active zone $\overline{\Omega}$ is indicated by the regions with the difference in bounds taking non-zeros (as per equation Eq. (\ref{eq:eq52})). It can be seen that increase in regularizing length scale has an effect on increase in portion of the active zone. Moreover, the active zone is concentrated close to the discontinuities. Figure \ref{fig:14}.d and \ref{fig:14}.e plots the Lipschitz projected field $\{d1,d2\}$ as per Eq. (\ref{eq:eq76}). $d_{tar}, d1$ and $d2$ along the midsection is plotted in Figure \ref{fig:14}.f. The regularizing lengths have the effect of smoothing the discontinuities with the slope being higher for lower regularizing length.

\section{Appendix D} \label{ap:d}
\begin{figure}
\centering
\captionsetup{justification=centering}
\begin{subfigure}{6cm}
\includegraphics[width= 6cm, height=6cm]{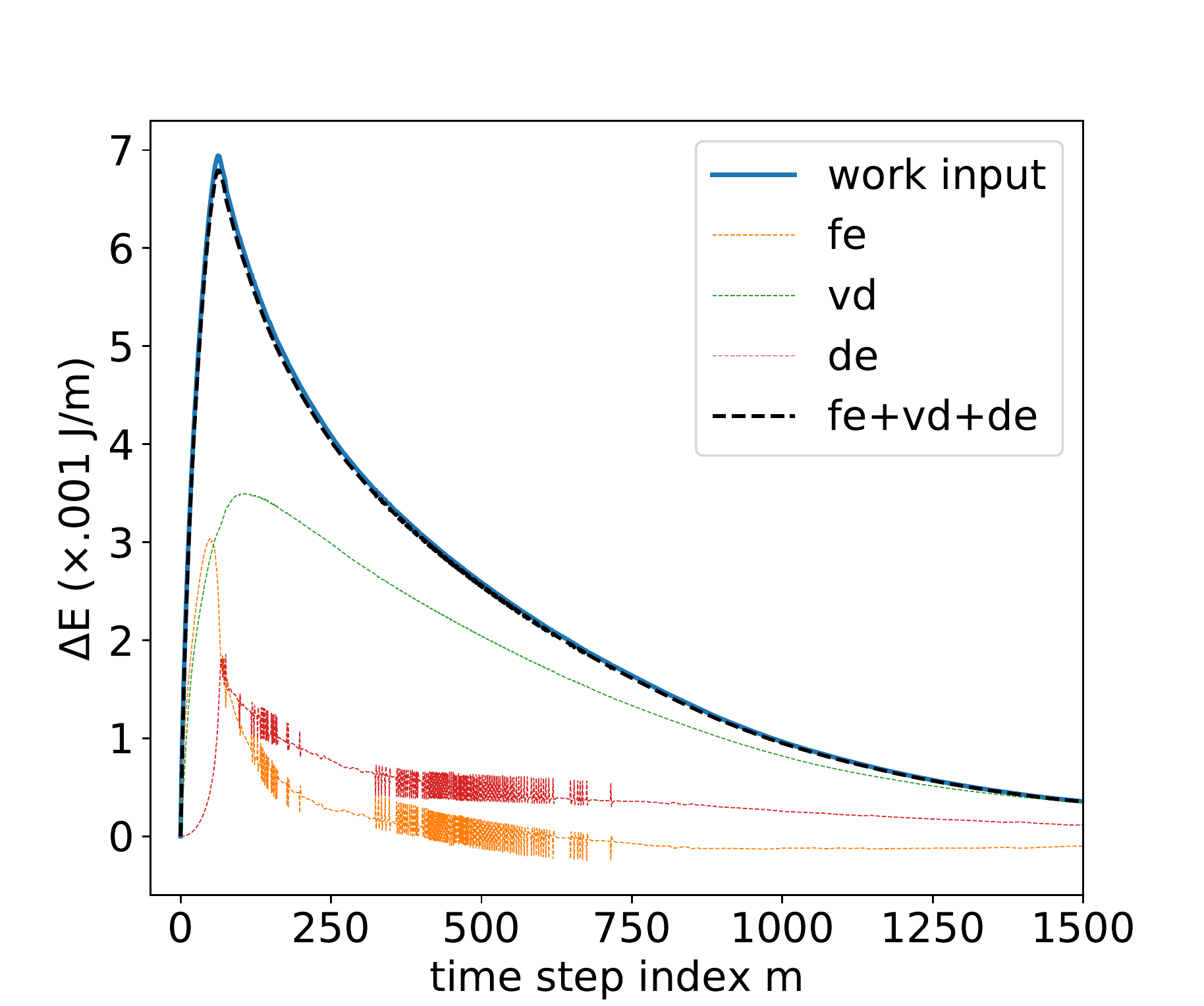}
   \caption{phase-field}
   \label{fig:15b}
\end{subfigure}
\begin{subfigure}{6cm}
\includegraphics[width= 6cm, height=6cm]{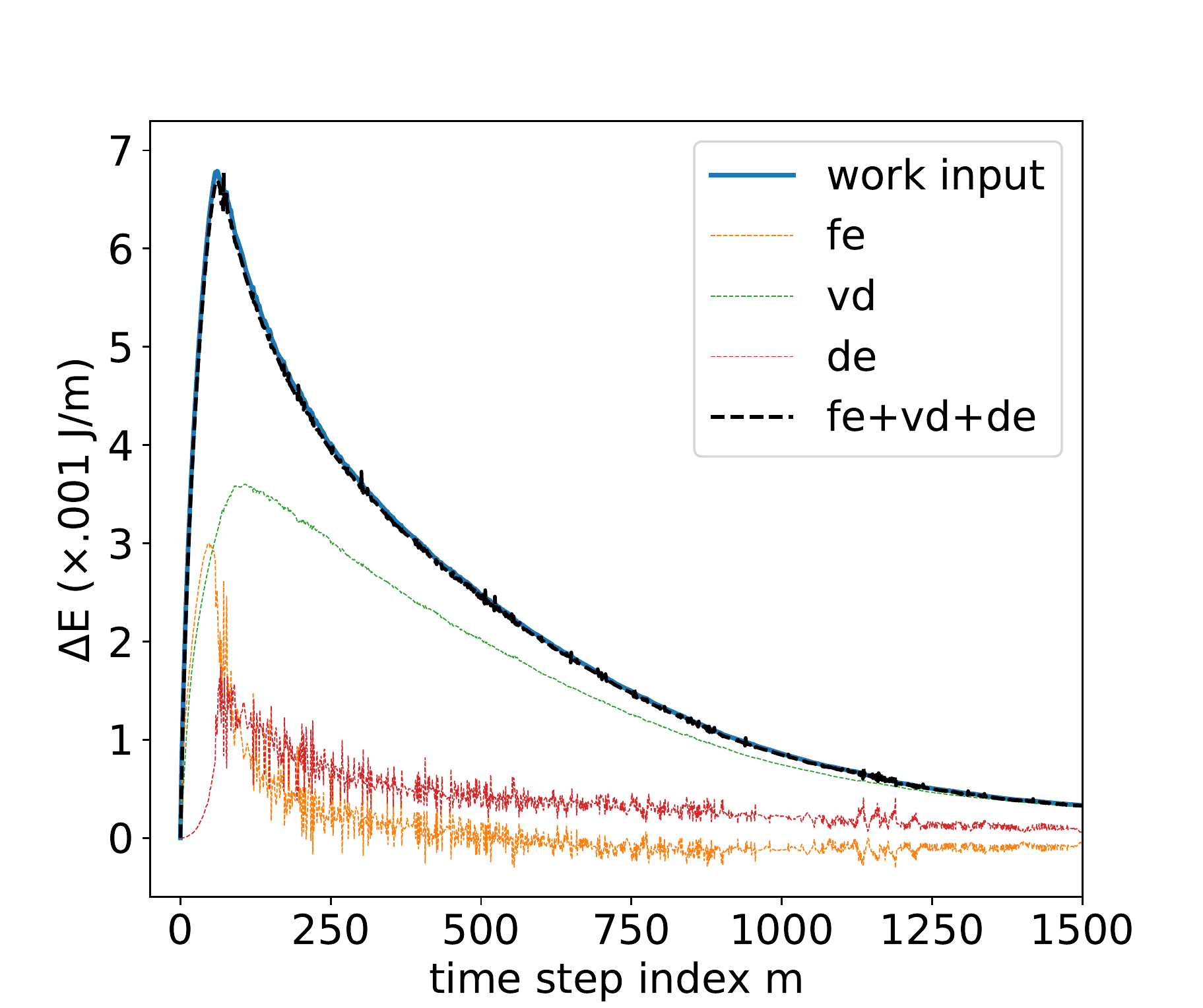}
   \caption{Lip-field }
   \label{fig:15a}
\end{subfigure}
\caption{ Incremental energies $E_{m+1}-E_m$ in the bulk as a function of time step index $m$. The free energy, viscous dissipation and damage energy in the bulk is denoted by \textit{fe,vd} and \textit{de}. Plots obtained for Test A at imposed displacement rate $v1$ }
\label{fig:15}
\end{figure}

In this section, the energy conservation of the developed model is illustrated. For this the incremental energy balance (energy between two consecutive time steps) for the Test A is considered. Figure \ref{fig:15} plots the incremental energy balance in the bulk for the TDCB geometry under the imposed loading rate $v2 = .1$ mm/s. $\Delta E$  corresponds to the time step index $m$ and represents the energy change $E^{m+1} -E^m$. The incremental work input is calculated as $R_y \Delta u_y$, where $\{R_y,\Delta u_y\}$ are the the observed reaction forces and the applied incremental displacements at time $t_{m+1}=t$.
It can be seen that both the approaches compare well in terms of energy.  The following observation could be made
\begin{itemize}
    \item Energy conservation in bulk is observed for both PF and LF with the applied incremental work used in a conservative manner to store free energy and dissipate viscous and damage energies.
    \item Viscous and damage dissipation are greater than zero satisfying second law of thermodynamics. In case of PF, though $\dot{d} \geq 0$ is not exactly guaranteed due to the use of a local history variable to drive damage, the damage energy in bulk is greater than zero. This indicates the use of history variable does not significantly affect the variational structure of the problem (given by Eq. (\ref{eq:eq40}))
    \item As viscous dissipation is not affected by damage, after the onset of damage growth, viscous dissipation remains the major consumer of the input work for the considered loading rate.
    \item The sudden rise in damage energy initially indicates faster rate of crack growth, but later the increase in damage energy starts to drop, indicating slow rate of crack growth.
    \item Moreover, relatively smooth profiles for damage energy is observed in case of PF because of the diffusive nature of phase-damage in the bulk. It is expected to get smoother damage energy profiles in case of LF by refining the mesh.
\end{itemize}

%

\end{document}